\documentclass[a4paper,fleqn,usenatbib]{mnras}

\usepackage[T1]{fontenc}
\usepackage{ae,aecompl}

\usepackage{amsmath}
\usepackage{amssymb}
\usepackage{graphicx}
\usepackage{hyperref}

\newcommand{\correction}[1]{{#1}}

\title[Protoplanetary disk ablation]{Hydrodynamic ablation of protoplanetary disks via supernovae}

\author[J. L. Close et al.]{J. L. Close$^{1}$\thanks{E-mail: py08jlc@leeds.ac.uk} and J. M. Pittard$^{1}$\\
$^{1}$School of Physics and Astronomy, University of Leeds, Leeds, LS2 9JT, UK}

\begin{document}

\date{\today}

\maketitle

\begin{abstract}
We present three-dimensional simulations of a protoplanetary disk subject to the effect of a nearby ($0.3\,$pc distant) supernova, using a time-dependent flow from a one dimensional numerical model of the supernova remnant (SNR), in addition to constant peak ram pressure simulations. Simulations are performed for a variety of disk masses and inclination angles. We find disk mass-loss rates that are typically $10^{-7}$ to $10^{-6}\,$M$_{\sun}\,$yr$^{-1}$ (but peak near $10^{-5}\,$M$_{\sun}\,$yr$^{-1}$ during the ``instantaneous'' stripping phase) and are sustained for around $200\,$yr. Inclination angle has little effect on the mass loss unless the disk is close to edge-on. Inclined disks also strip asymmetrically with the trailing edge ablating more easily. Since the interaction lasts less than one outer rotation period, there is not enough time for the disk to restore its symmetry, leaving the disk asymmetrical after the flow has passed. Of the low-mass disks considered, only the edge-on disk is able to survive interaction with the SNR (with 50\% of its initial mass remaining). At the end of the simulations, disks that survive contain fractional masses of SN material up to $5 \times 10^{-6}$. This is too low to explain the abundance of short-lived radionuclides in the early solar system, but a larger disk and the inclusion of radiative cooling might allow the disk to capture a higher fraction of SN material.
\end{abstract}


%__________________________________________________________________

\section{Introduction}

It is widely believed that massive stars can trigger the formation of lower mass stars in their surrounding regions, leading to frequent associations of massive stars, and young low mass stars. For instance, there are $\sim2000$ low-mass stars within 2pc of the centre of the Trapezium cluster \citep{1998ApJ...492..540H}, and stars with circumstellar disks \citep{1993ApJ...410..696O, 1996AJ....111.1977M, 1998AJ....116..293B} within a few tenths of a parsec of the central star, $\theta^1$ Ori C. As massive stars are often the source of strong winds, high ionising radiation flux, and eventual supernovae, these regions are potentially hostile to protoplanetary disks and the subsequent formation of planets. The interactions massive stars have on other stars and their disks can be broadly categorised into gravitational, radiation or ablation.

When two stars pass one another, the associated disks become perturbed by the gravitational interaction. If this interaction is strong enough, material from the disk can become unbound \citep{1993MNRAS.261..190C}. \cite{2001MNRAS.325..449S} use N-body simulations to model the dynamics of the stars in the Orion Nebula. They find that only a small fraction of the stars interact closely on the expected disk dispersal timescales, and so close encounters are not a major contributor to protoplanetary disk dispersal.

Direct observational evidence of the effects of radiation on disks is seen in the form of the proplyds in the Orion Nebula (e.g. \citet{1994ApJ...436..194O, 2001IAUS..205..236M}). The radiation from $\theta^1$ Ori C causes a photoevaporation flow from the disks which collides with the oncoming stellar wind. This results in a number of cometary-shaped ionization fronts associated with stars around $\theta^1$ Ori C. Estimates put the mass-loss rate of the proplyds at $\sim4 \times 10^{-7} M_{\sun}\,$yr$^{-1}$ \citep{1999AJ....118.2350H}. If sustained over the lifetime of the cluster this would mean that the stars would have had unrealistically massive disks (greater than the star's mass) at some point in the past. However, $\theta^1$ Ori C may have switched on only $\sim10^{5}\,$years ago \citep{2001MNRAS.325..449S}, in which case it is unlikely that the combination of photoevaporation and viscous accretion will have significantly depleted the disks \citep{2011ppcd.book.....G}. This suggests that the Orion Nebula is in a relativity short-lived stage of its evolution, which is consistent with the lack of proplyd-like objects in other clusters (e.g. \citet{1997ASPC..119..131S, 2006ApJ...650L..83B}).

\correction{
There is now a lot of observational evidence that massive stars influence the disks of neighbouring stars. Cometary tails are seen to extend from protoplanetary disks in Hubble Space Telescope images of the Orion Nebula \citep[e.g.][]{2005AJ....129..382S, 2008AJ....136.2136R}, while sub-millimeter observations of the Trapezium Cluster reveal that  disks within a few tenths of a parsec of $\theta^1$ Ori C are truncated \citep{2009ApJ...694L..36M, 2010ApJ...725..430M, 2014ApJ...784...82M}. However, the degree to which massive stars affect nearby disks remains uncertain. For instance, \citet{2015ApJ...802...77M} claim that there is no evidence for disk truncation in the Flame Nebula (NGC 2024), though it is somewhat younger and hosts a less massive star compared to the Trapezium cluster. In addition, \citet{2015ApJ...811...10R} argue that there is no evidence for the depletion of the inner disks around pre-main-sequence stars in the vicinity of O-type stars, even very luminous O2-O5 stars. In contrast, \citet{2016ApJ...825..125P} claim that rapid disk evolution on $<1\,$Myr timescales has produced a significant population of intermediate-mass pre-main-sequence stars lacking inner dust disks.
}

\correction{
Likewise, significant differences in the mass-loss rate and lifetime of disks exist in theoretical studies of their photoevaporation. Some studies claim lifetimes of only $10^5\,$yrs \citep[e.g.][]{1998ApJ...499..758J, 1999ApJ...515..669S}, substantially shorter than the $10^6\,$-$\,10^7\,$yrs lifetime of disks subject only to viscous accretion \citep[and references therein]{2001ApJ...553L.153H, 2007ApJ...662.1067H}. In contrast, a recent model combining viscous accretion and external photoevaporation predicts that disk lifetimes are shortened by only a factor of a few \citep{2013ApJ...774....9A}. In addition, if the outer parts of the disk can be totally removed, the inner disk may rapidly disappear if it is starved of material needed to offset accretion onto the star \citep{2013ApJ...774....9A}. Thus the significance of external photoevaporation on disk lifetimes remains highly uncertain.
}

Supernova remnants (SNRs) provide another source of agitation for protoplanetary disks. There is evidence in $\pi$~Sco of circumstellar disks similar to those in the Orion nebula \citep{1992ApJ...388..495B} and that a supernova occurred nearby $\sim 10^{6}\,$yrs ago \citep{1992A&A...262..258D}. \cite{2003ARA&A..41...57L} compiled a catalog of embedded clusters and find that the vast majority ($70$ - $90\%$) of stars form in clusters of $> 100$ members, and the majority of those ($\sim 75\%$) are in clusters containing stars massive enough to give rise to supernovae \citep{2005ASPC..341..107H}.

Analytic estimates for the ablation of protoplanetary disks due to a SNR were presented by \citet{2000ApJ...538L.151C}. He found that for typical disk and supernova parameters, partial stripping of the disk can occur, but typically not its complete disruption. This suggests that although the disks are affected, they can survive such events.

The interaction of a SNR with a protoplanetary disk is also of interest from the point of view of injecting short-lived radionuclides (SLRs) into the early solar system (e.g. \citet{1977Icar...30..447C, 2005ASPC..341..548J, 2006ApJ...652.1755L}). Analyses of meteorites suggest that the early solar system had a higher abundance of SLRs than the average interstellar medium (ISM) value \correction{\citep[see e.g.][for $^{26}Al$]{2006Natur.439...45D}}. This means the SLRs must have been injected into the solar system at some point during its history, and due to the short half-life of $^{60}$Fe \correction{\citep [$2.6\,$Myr,]{2009PhRvL.103g2502R})} this constrains the injection event to a time close to the solar system's formation. Given that supernovae are abundant sources of nucleosynthesis and they are often in close proximity to young, low mass stars, they are seen as a good candidate for the source of SLRs in the early solar system. While some SLRs can be explained by production internal to the solar system, $^{60}$Fe is practically impossible to produce locally \citep{1998ApJ...506..898L, 2006ApJ...640.1163G}.

If the formation of the solar system was triggered by a SNR this could provide a source of SLRs (see e.g. \cite{1996ApJ...468..784F, 2010ApJ...717L...1B, 2012ApJ...756L...9B, 2014ApJ...788...20B, 2015ApJ...809..103B}). While this is a plausible model, it puts extra constraints on the timing of the formation of the disk. This is estimated to take $\sim 0.1 - 10\,$Myr \citep[see e.g.][]{1993prpl.conf....3S, 1998ApJ...508..291V,2004ApJ...616..283T}. This is on top of the time needed to form small bodies in the solar system, which is $\sim 1 - 10\,$Myr \citep{2005ASPC..341..131H}. If SLRs are injected during the molecular cloud stage, disk formation must occur quickly such that enough SLRs exist in the early solar system. 

In the current paper we investigate injection into an already formed disk as an alternative to this model. Injection would have to occur in the relativity early stages of the disk's lifetime, before small bodies begin to form. \correction{It is currently unknown at what point material can be injected and homogenised in order to be captured by forming meteorites.} As all these processes are likely to occur simultaneously, simulations that track both injection and the subsequent formation into meteorites would go a long way to answering this question. However, this is beyond the scope of the current paper.

The abundance of SLRs is typically parametrised by the ratios of $^{60}$Fe and $^{26}$Al to their stable isotopes. \cite{2005LPI....36.1827Q} and \cite{2006ApJ...639L..87T} find a $^{60}$Fe/$^{56}$Fe ratio of $\sim 3 - 7 \times 10^{-7}$. However, the technique used (secondary ion mass spectrometry (SIMS)) has been shown to give a significant positive bias \citep{2011NIMPB.269.1910O}. Subsequent studies using a different technique \citep{2012E&PSL.359..248T} show a lower ratio of $5 \times 10^{-8}$, although \cite{2014GeCoA.132..440M} use an improved SIMS method and find a ratio of $7 \times 10^{-7}$. Therefore, the true ratio is still unclear. However, if the lower estimates for the amount of $^{60}$Fe are correct, then the $^{60}$Fe/$^{26}$Al ratio becomes much lower than that generated by supernovae \citep{2012A&A...545A...4G}. A possible resolution to this problem is suggested by \cite{2016MNRAS.462.2777G}, who note that different sized dust grains are injected at different rates and if they preferentially carry different SLRs this could have a significant effect on the $^{60}$Fe/$^{26}$Al ratio.

The amount of $^{60}$Fe produced by a supernova varies depending on the mass and metallicity of the exploding star. Calculations by \cite{1995ApJS..101..181W} put the $^{60}$Fe mass fraction at $\sim 2 \times 10^{-6}$ for a $20 M_{\sun}$ supernova. Using initial solar system abundances from \cite{2003ApJ...591.1220L}, \cite{2006ApJ...652.1755L} calculate a meteoritic $^{60}$Fe mass fraction of $1.10 \pm 0.38 \times 10^{-9}$. 

We can define an enrichment fraction, $f_{enr}$, as the mass of supernova material mixed into the disk as a fraction of the disk mass. Then the change in mass fraction, $X$, of a given species, $i$, after enrichment can be calculated as:
\begin{equation}
	X_{i, after} = \left(1 - f_{enr} \right) X_{i, before} + f_{enr} X_{i, supernova}.
\end{equation}
If the initial solar system has no significant $^{60}$Fe, then $f_{enr}$ becomes simply:
\begin{equation}
	f_{enr} = \frac{X_{i, after}}{X_{i, supernova}}.
\end{equation}
Using $X_{i, after} = 1.10 \times 10^{-9}$ and $X_{i, supernova} = 2 \times 10^{-6}$ from above this gives $f_{enr} = 5.5 \times 10^{-4}$. Doing the same calculation for $^{26}$Al ($X_{i, after} = 3.77 \times 10^{-9}$ and $X_{i, supernova} = 3 \times 10^{-6}$) gives $f_{enr} = 1.3 \times 10^{-3}$.

\correction{
The assumption of zero $X_{i, before}$ is a contested one. It has been suggested \citep[e.g.][]{2009ApJ...696.1854G, 2016ApJ...826..129Y} that successive supernovae and massive stellar winds can allow a molecular cloud to self-enrich, such that stellar systems form with enriched material already embedded. This scenario is explored in simulations by \citet{2013ApJ...769L...8V} and \citet{2016ApJ...826...22K}. Using a turbulent periodic box, the evolution of a massive ($\sim 10^5 M_{\sun}$) star forming region is followed over $20\,$Myr. It is found that supernova feedback enriches the star forming gas to a level that is largely consistent with that of the early solar system, although the authors rely on numerical diffusion to mix the SN ejecta and cold gas. This model also remains somewhat controversial as it requires an age spread of star formation within the same cloud, which is at odds with observations \citep{2000ApJ...530..277E, 2011MNRAS.418.1948J, 2014prpl.conf..219S} and simulations \citep{2003MNRAS.343..413B, 2012MNRAS.424..377D}. 
%Other work examining clusters of $10^{3}-10^{4}$ stars finds that the pollution of protoplanetary discs by supernova ejecta can vary considerably within the cluster: many discs had negligible enrichment, but 10-30\% were enriched, typically to levels comparable to or greater than the solar system \citep{2016MNRAS.462.3979L}. However, these works necessarily adopt simplified prescriptions for the pollution mechanism, and make assumptions about the isotopic mixing and protoplanetary disc lifetimes.
%2016MNRAS.462.3979L = Lichtenberg, Parker \& Meyer 2016 MN, 462, 3979
While a non zero value for $X_{i, before}$ doesn't necessarily change the scenario investigated in this work, it would increase the mass fractions we see.
}

Variations in the supernova mass can also produce quite large differences in the mass fraction, not to mention the inherent uncertainties in supernova modelling. \correction{For example, \citet{2006ApJ...647..483L} calculate a $^{60}$Fe mass fraction for a $20 M_{\sun}$ supernova at $7.8 \times 10^{-7}$, $2.5\times$ less than that calculated by \cite{1995ApJS..101..181W}. Their $^{60}$Fe mass fractions for the next nearest supernova masses ($17 M_{\sun}$ and $25 M_{\sun}$) vary from the $20 M_{\sun}$ mass fraction by a factor of $\sim 1.5$.} With this in mind $f_{enr}$ is only \correction{constrained} to a factor of two or three at best.

Hydrodynamical simulations of the interaction of a SNR with a circumstellar disk were performed by \cite{2007ApJ...662.1268O}. They found that only a small fraction ($\sim 1\%$) of the disk mass is removed, in contrast to the $\sim 13\%$ expected from the analytical prediction of \cite{2000ApJ...538L.151C}. They attribute this to the cushioning and deflecting effect of the bow shock in addition to the compression of the disk further into the gravitational well of the central star. However their simulations are limited to two dimensions and face-on impacts. While an inclined disk might be expected to be disadvantageous for ablation, the geometry of the bowshock that shields the disk is likely to be substantially different and the impacting flow acts with the rotation of the disk on one side, reducing the threshold for ablation. This scenario therefore needs to be investigated. Ouellette et al. also found that only $\sim 1\%$ of the passing ejecta is captured by the disk. This is not enough to explain the observed abundances, although they suggest that dust grains might be a more efficient mechanism for injecting SLRs into the disk.

To investigate this possibility, \cite{2010ApJ...711..597O} incorporated dust grains into their simulations. They found that about $70\%$ of dust grains larger than $0.4\,\mu\,$m are injected into the disk. This could potentially be enough to explain the observed abundances, although it relies on higher than observed dust condensation in the ejecta and the ejecta to be clumpy with the disk being hit by a high density clump of ejecta. They estimate the probability of these conditions being met at $0.1 - 1\%$ and that the solar system may be atypical in this respect.

The mixing effects of clumpy supernova ejecta interacting with a molecular cloud were investigated by \cite{2012ApJ...756..102P}. When the mixing is efficient the enrichment fraction, defined as the ratio of ejecta mass to cloud mass, is $\sim 10^{-4}$. This is approximately the correct level, although as the SLRs are injected into the molecular cloud, and it takes of order a Myr to form stars and disks, the abundance of $^{60}$Fe could be significantly reduced in this time.

The formation and subsequent evolution of a star-disk system under the effects of a supersonic wind were investigated by \cite{2014MNRAS.444.2884L}. They found that a disk can indeed form and survive in these conditions: a $10^{-3}M_{\sun}$ disk remains after being exposed to the wind for $0.7\,$Myr. However, their disk radius after formation is $\sim1000\,$au, whereas disks near massive stars are typically photoevaporated down to tens of au \citep{1998ApJ...499..758J}. The flow speed is also much slower than that adopted by \cite{2007ApJ...662.1268O}, which in turn reduces the ram pressure significantly. Therefore, while their work was useful for investigating triggered star formation, the region of parameter space that they explored is unrealistic for disk ablation via nearby supernova.

Recently \cite{2016MNRAS.462.2777G} performed three dimensional simulations of the interaction of a SNR with a large ($8.8\,$pc), clumpy molecular cloud, including dust grains. They found that the majority of large dust grains are injected into the molecular cloud, within $0.1\,$Myr of the supernova explosion. They note that if $^{60}$Fe and $^{26}$Al preferentially condense onto different sized grains this could explain the discrepancy in the ratio between the two SLRs.

This paper aims to improve and expand on the 2D hydrodynamical calculations of \cite{2007ApJ...662.1268O} in a number of ways. Firstly, by doing fully 3d simulations, more complex dynamics can be studied, and the angle between the disk and the flow can be varied. Secondly, we consider a range of disk masses ($1\times$, $0.1\times$ and $0.01\times$ their canonical disk mass) as disk density is the main factor in determining the extent to which a disk is ablated. By considering a number of disk masses spanning a range of midplane densities, we can investigate how different disks react and at what point the disk becomes significantly affected. Finally, we adopt time-dependent flow properties from a 1D simulation of the SNR. This provides more accurate flow conditions past the disk, compared to the analytical approximation used by \cite{2007ApJ...662.1268O}, particularly in the early stages of the SNR's expansion. We aim to better determine the nature of the interaction, the mass-stripping rate of the disk, and the injection rate of SLRs into the disk.

%__________________________________________________________________

\section{Models}
\label{sec:sim}

\subsection{Overview}

A stellar disk is simulated in three dimensions, with mass injected onto the grid to simulate the impact of a SNR. The strength of the flow is dependent on the size and distance of the supernova and is time dependent. To calculate this precisely, a simple, one dimensional simulation of a spherical SNR is performed. The evolution of the density, velocity and temperature are recorded at the desired distance from the supernova, which can then be used by the main simulation to control the properties of the wind. This is in many ways simpler than performing an analytical approximation (e.g. \citet{1999ApJ...510..379M}), and has the advantage of reproducing all the features of a realistic SNR, particularly in the early expansion.

\subsection{The supernova remnant}
\label{sec:snr}

\begin{table}
	\caption[SNR parameters]{SNR parameters}
	\label{table5:snr}
	\centering
	\begin{tabular}{l c}
		\hline
		Variable & Value \\
		\hline
   		$M_{ej}$ & $20\,$M$_{\sun}$ \\
   		$E_{ej}$ & $10^{44}\,$J \\
   		$\rho_{amb}$ & $2.34 \times 10^{-24}\,$g$\,$cm$^{-3}$ \\
   		$T_{amb}$ & $10^{4}\,$K \\
		\hline
	\end{tabular}
\end{table}

The supernova has an ejecta mass of $20\,$M$_{\sun}$ and an ejecta energy of $10^{51}\,$erg (following \cite{2007ApJ...662.1268O} to represent a typical type II supernova), and is initially confined to a radius of $2000\,$au. The ambient medium is set to a density of $2.34 \times 10^{-24}\,$g$\,$cm$^{-3}$ and a temperature of $10^4\,$K. The simulation parameters are noted in Table \ref{table5:snr}. For all simulations, the mean molecular weight, $\mu = 2.4$, and $\gamma = 5 / 3$. The supernova calculations were performed in spherical symmetry on a uniform one-dimensional grid, extending to $r = 10^{5}\,$au with 3200 cells. The same calculations were done at half and quarter resolution with identical results.

\begin{figure*}
	\centering
	\begin{tabular}{@{}c@{}}
		\includegraphics[width=0.5\textwidth]{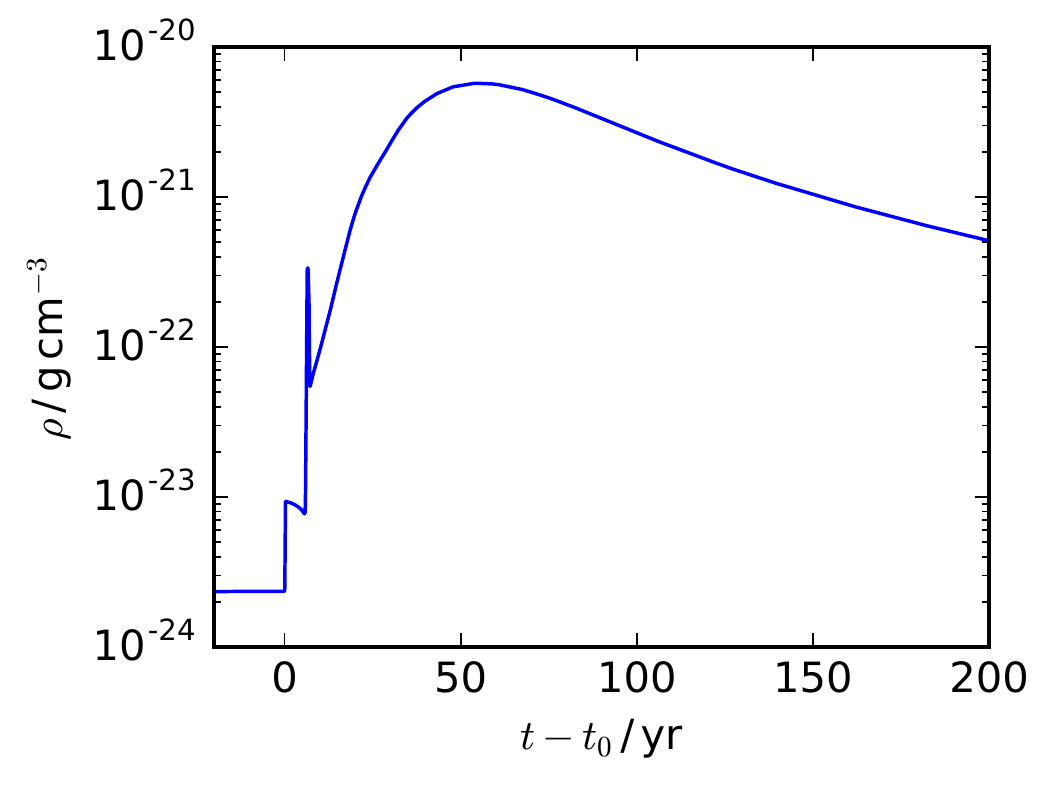}
		\includegraphics[width=0.5\textwidth]{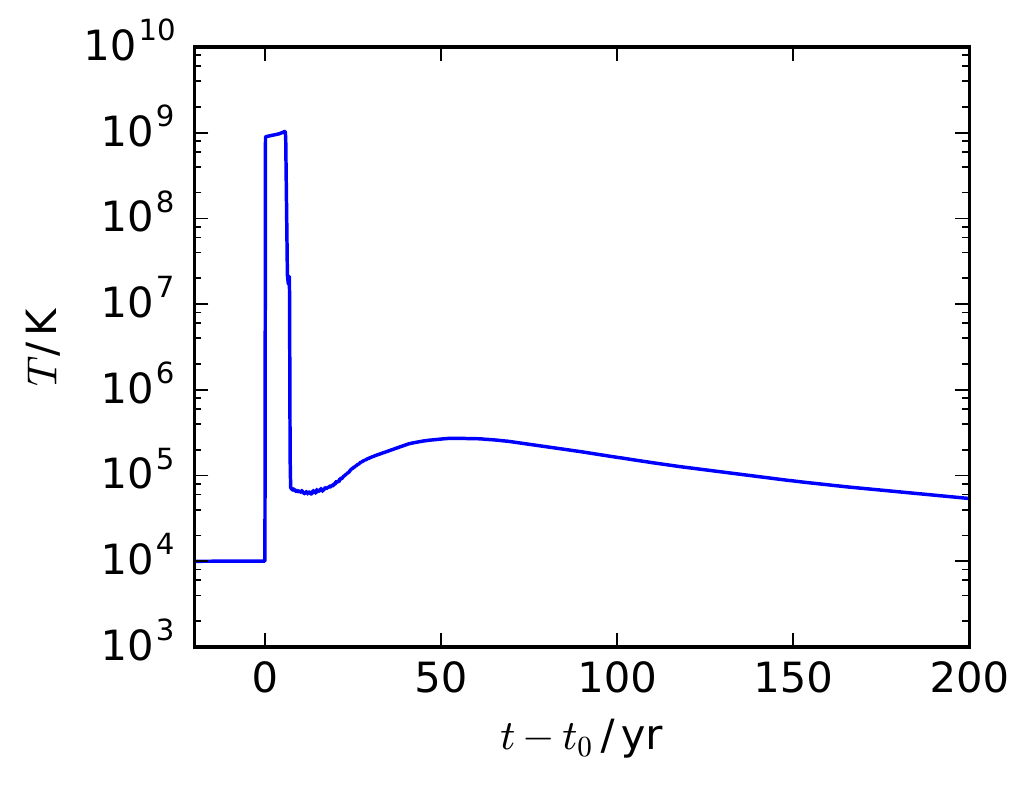} \\
		\includegraphics[width=0.5\textwidth]{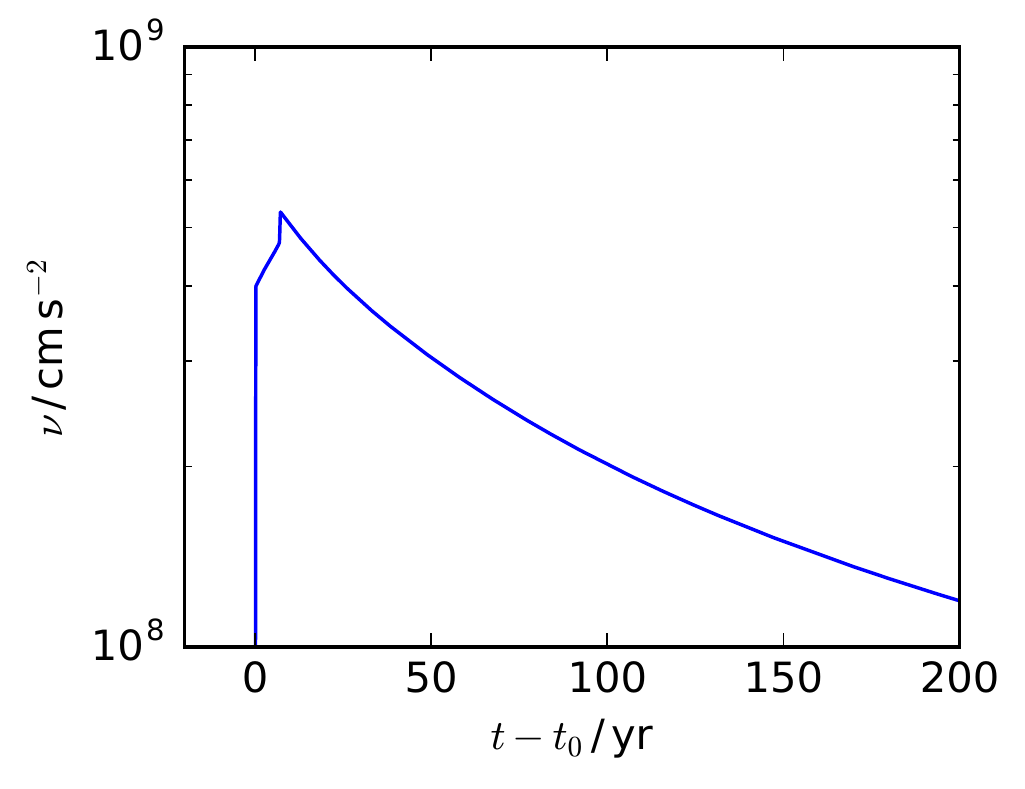}
	\end{tabular}
	\caption[Density, temperature and velocity of the fluid $0.3\,$pc away from the supernova.]{Density, temperature and velocity of the fluid $0.3\,$pc away from the supernova as described in Section \ref{sec:snr}. $t_{0}$ is the time at which the SNR shockwave first reaches a radius of $0.3\,$pc ($t_{0} \approx 55$\,yr).}
	\label{fig5:snr}
\end{figure*}

\begin{figure}
	\centering
	\includegraphics[width=0.5\textwidth]{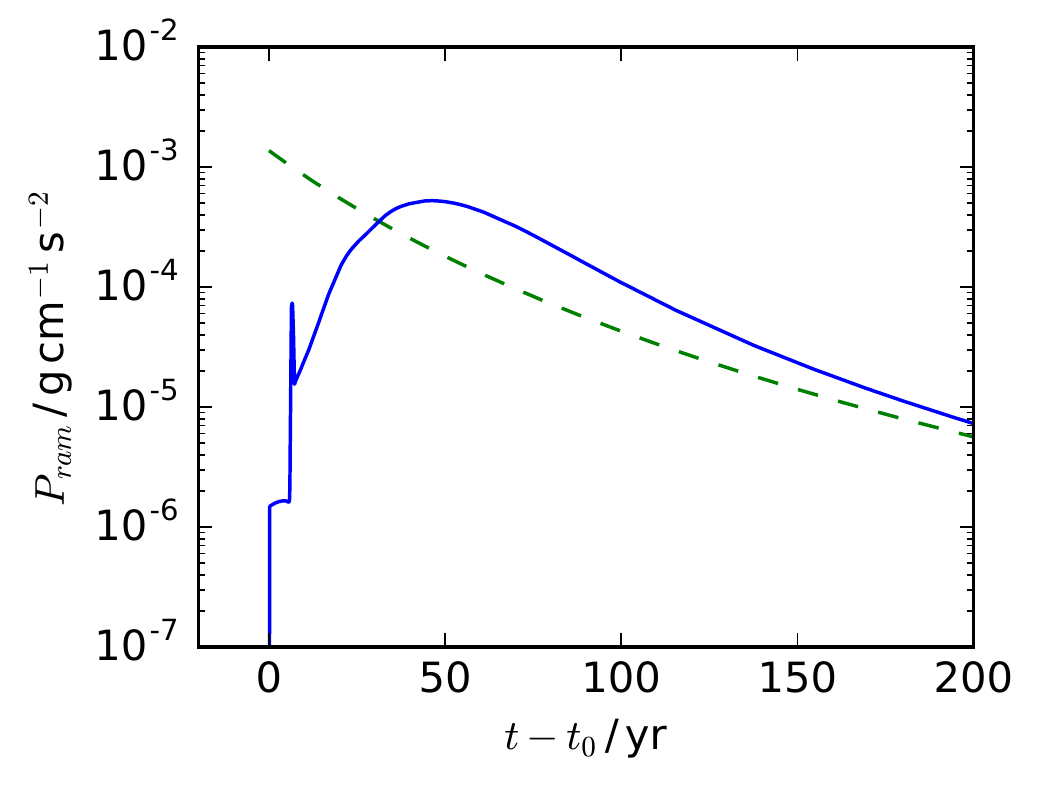}
	\caption[Ram pressure $0.3\,$pc away from the supernova.]{Ram pressure $0.3\,$pc away from the supernova as described in Section \ref{sec:snr}. $t_{0}$ is the time at which the SNR shockwave first reaches $0.3\,$pc. The dashed green line shows the ram pressure used by \citet{2007ApJ...662.1268O} for comparison.}
	\label{fig5:snr_ram}
\end{figure}

Figure \ref{fig5:snr} shows how the fluid variables change over time at a point $0.3\,$pc from the origin of the SNR. $0.3\,$pc is chosen to aid comparison with \cite{2007ApJ...662.1268O} and is consistent with observations of the Orion Trapezium Cluster. These values will determine the properties of the flow past the circumstellar disk in the 3d simulation. The supernova ejecta takes $55.4\,$yr to reach $0.3\,$pc, which is insignificant compared to the half-lives of the SLRs so there is no significant decay during the travel time. The SNR is still in its free expansion stage at this point. The strength of the flow is parametrised by the ram pressure, $P_{ram} = \rho \nu^2$, and is shown in Figure \ref{fig5:snr_ram}. Also shown for comparison is the analytical ram pressure curve used by \citet{2007ApJ...662.1268O}, which is only a close approximation to the numerical results at late times. The interaction is strongest in the early stages, where the approximation is poor.

\subsection{The circumstellar disk}
\label{sec:disk}

\begin{table}
	\caption[Disk model parameters]{Disk model parameters}
	\label{table5:disk}
	\centering
	\begin{tabular}{l c}
		\hline
		Variable & Value(s) \\
		\hline
		$R_{0}$ & $1\,$au \\
		$T_{0}$ & $400\,$K \\
		$\rho_{0}$ & $\left\{\begin{tabular}{@{\ }l@{}}
			$3.5 \times 10^{-13}\,$g$\,$cm$^{-3}$ \\ $3.5 \times 10^{-12}\,$g$\,$cm$^{-3}$ \\ $3.5 \times 10^{-11}\,$g$\,$cm$^{-3}$
		\end{tabular}\right.$ \\
		$M_{*}$ & $1\,$M$_{\sun}$ \\
		\hline
	\end{tabular}
\end{table}

The disk model is adapted from \cite{2013ApJ...779...59G}. The temperature is defined to be constant across z, and to be inversely proportional to cylindrical radius, R,
\begin{equation}
	T(R) = T_{0} (R / R_{0})^{-1}.
	\label{eqn5:temp}
\end{equation}
This gives the disk a constant opening angle. A constant opening angle is typically desirable at it allows the disk edge to align with cell boundaries in a spherical-polar domain. While this is not a concern for these simulations, it aids comparison with works in the literature. Defining the midplane density as
\begin{equation}
	\rho_{mid}(R) = \rho_{0} (R / R_{0})^{-3/2},
	\label{eqn5:rhomid}
\end{equation}
and enforcing hydrostatic equilibrium vertically defines the three dimensional density structure, which can be derived as
\begin{equation}
	\rho(\bmath{r}) = \rho_{mid} \, \mathrm{exp}\left(\frac{GM_{*}}{c_{s}^{2}}\left[\frac{1}{r}-\frac{1}{R}\right]\right),
	\label{eqn5:rhofull}
\end{equation}
where $c_{s}$ is the isothermal sound speed, and $r$ is the distance to the centre of the disk. Finally, the angular velocity, $\omega$, is set to achieve radial equilibrium,
\begin{equation}
	\Omega(\bmath{r}) = \Omega_{k}(R) \sqrt{\frac{R}{r} - \frac{5}{2}\left(\frac{c_{s}}{\Omega_{k}(R)R}\right)^2},
\end{equation}
where $\Omega_{k}(R) = \sqrt{GM_{*}}R^{-3/2}$ is the Keplerian angular velocity. The free disk parameters are chosen such that the disk resembles that described in \cite{2005ASPC..341..527O}. As they observe no significant ablation of their disk, lower disk mass simulations are also performed, which are likely to be more susceptible to ablation. The values are summarised in Table \ref{table5:disk}. The disk is truncated at an inner radius of $4\,$au and an outer radius of $40\,$au. The temperature of the disk ranges from $100\,$K to $10\,$K from the inner to outer disk boundaries. The outer rotation period is $\sim250\,$yr and the inner rotation period is $\sim10\,$yr. The initial ambient medium ($\rho_{amb}$ and $T_{amb}$) is set to the same as that of the supernova simulation for consistency (see Table \ref{table5:snr}). The disk was evolved in isolation and found to be stable for several outer rotation periods.

It should be noted that much bigger disks (with radii up to $\sim1000\,$au see e.g. \cite{2015ApJ...808...69B}) have been observed. Such disks have a much larger surface area and the outer regions are less strongly held by the gravitational field, meaning they will lose mass to ablation more readily. However, a $40\,$au disk is used here to aid comparison with previous work. Without good statistics of the disk radius at formation it is also hard to say what is more typical and in any case it likely depends on unknown quantities such as the conditions under which the cluster formed. At larger radii, both the gravitational field strength and the disk's surface density decrease, making the material more tenuously held and more easily ablated. For a given disk model, only the central density and stellar mass determine the stripping radius: the stripping radius does not depend on the initial disk radius. With this in mind, some insight into the behaviour of larger disks can also be gleaned from our simulations.

%densities of $4.38 \times 10^{-11}\,$kg$\,$m$^{-3}$ and $1.38 \times 10^{-12}\,$kg$\,$m$^{-3}$

%__________________________________________________________________

\section{The simulations}

\begin{table}
	\caption[Summary of simulation parameters.]{Summary of the simulations. The disk angle ($i$) is defined as the difference in angle between the angular momentum vector of the disk and the vector of the flow direction, making $0^{\circ}$ face-on and $90^{\circ}$ edge-on. $M_J$ is the mass of Jupiter.}
	\label{table5:sims}
	\centering
	\begin{tabular}{c c c c c}
		\hline
		Simulation & $\rho_{0}\,$/$\,$g$\,$cm$^{-3}$ & $M_{disk}\,$/$\,M_{J}$ & $i$ & Dynamic flow? \\
		\hline
		\textbf{const00high} & $3.5 \times 10^{-11} $ & 8.22   & $0^{\circ}$  & n\\
		\textbf{const00med}  & $3.5 \times 10^{-12} $ & 0.822  & $0^{\circ}$  & n\\
 		\textbf{const00low}  & $3.5 \times 10^{-13}$ & 0.0822 & $0^{\circ}$  & n\\
 		\textbf{const45low}  & $3.5 \times 10^{-13}$ & 0.0822 & $45^{\circ}$ & n\\
 		\textbf{const90low}  & $3.5 \times 10^{-13}$ & 0.0822 & $90^{\circ}$ & n\\
 		\textbf{dyn00low}    & $3.5 \times 10^{-13}$ & 0.0822 & $0^{\circ}$  & y\\
 		\textbf{dyn45low}    & $3.5 \times 10^{-13}$ & 0.0822 & $45^{\circ}$ & y\\
 		\textbf{dyn90low}    & $3.5 \times 10^{-13}$ & 0.0822 & $90^{\circ}$ & y\\
		\hline
	\end{tabular}
\end{table}

The calculations were performed with the hierarchical adaptive mesh refinement (AMR) code, \texttt{MG} \citep{1991MNRAS.250..581F}. The code uses the Godunov method, solving a Riemann problem at each cell interface, using piece-wise linear cell interpolation and MPI-parallelization. The Eulerian equations of hydrodynamics are solved using the second order upwind scheme described in \citet{1991MNRAS.250..581F}. Refinement is on a cell-by-cell basis and is controlled by error estimates based on the difference between solutions on different grids, i.e. the difference between the solutions on $G_{n-1}$ and $G_n$ determine refinement to $G_{n+1}$. Spatial resolution is doubled on each refinement level. Further details on the AMR method can be found in \cite{ISI:000226971600016}.

All simulations are performed on a three dimensional Cartesian grid, with the disk situated at the origin, and the plane of rotation aligned with the X-Y plane. For each simulation a ``flow injection region'' is defined where the values of the grid cells are set to the current SNR flow properties at the beginning of each time step. This region is placed $96\,$au from the origin. The simulation region extends to $\pm256\,$au in all directions, except in the cases where this would create a large flow injection region. For all simulations the lowest grid level has a resolution of $8\,$au, with 4 additional grid levels giving an effective resolution of $0.5\,$au. This gives $80$ cells per disk radius. Detailed convergence studies for shock-cloud interactions \citep{2016MNRAS.457.4470P} indicate that a resolution of 32-64 cells per cloud radius is needed for signs of convergence, including in the mixing fraction. Although the scenarios are different, this provides some confidence that our simulations are of adequate resolution. They are also at a higher resolution than any previous studies of interactions of this type. A detailed resolution study is left to future work.

There are two reasons the disk is kept in the same orientation and the wind direction is changed (as opposed to vice versa). Firstly, if the plane of rotation is not aligned to the direction of the grid, instabilities can develop and cause the disk to fragment with no outside influence (e.g., \citet{1984JCoPh..56...65D, 2010MNRAS.405..274H, 2015MNRAS.450...53H}). Secondly, the grid is split among processors by dividing the domain along one of the grid directions. Placing the disk such that the divides split the disk across processors helps to distribute the computational load more evenly and allows the simulations to run more efficiently.

A set of eight simulations are performed. Table \ref{table5:sims} details the differences between them. Some simulations are done with a constant flow, as this is the most straight forward to compare against analytical approximations. The parameters of the constant flow are defined by the peak ram pressure point of the SNR (see Figures \ref{fig5:snr} and \ref{fig5:snr_ram}) and occurs $55.8\,$yrs after the disk is first hit by the SNR. The peak is taken as it represents the worst case scenario for the survival of the disk. Specifically this is a density of $5.74 \times 10^{-21}\,$g$\,$cm$^{-3}$, a velocity of $2.87 \times 10^{8}\,$cm$\,$s$^{-1}$ and a temperature of $2.72 \times 10^{5}\,$K. The corresponding ram pressure is $4.74 \times 10^{-4}\,$g$\,$cm$^{-1}\,$s$^{-2}$. In contrast the flow in the dynamic simulations follows the temperature, density and velocity of the calculated SNR, shown in Figure \ref{fig5:snr}.

\subsection{Defined quantities}
\label{sec5:def}
In order to quantify the mass of the disk and the amount of SN ejecta that is captured by the disk, we define several quantities. The mass of the disk, $M_{disk}$, is defined as the total mass of gas that is bound to the central star's gravitational field, i.e. the velocity of gas in a given cell is less than the escape velocity at that point in space. The simulation uses an advected scalar to track the SLR enriched SN ejecta. Disk material is given a scalar value of $0.0$ and the enriched SN ejecta is given a scalar value of $1.0$. Thus, for a given grid cell, the product of the advected scalar, the cell density and the cell volume gives the mass of enriched material in that cell. We define the mass of captured enriched material, $M_{enr}$, as the total mass of enriched material that is gravitationally bound. Thus the enrichment fraction $f_{enr} = M_{enr} / M_{disk}$. 

%__________________________________________________________________

\section{Analytical Approximations}
\label{sec5:analytic}

\begin{figure}
	\centering
	\includegraphics[width=0.5\textwidth]{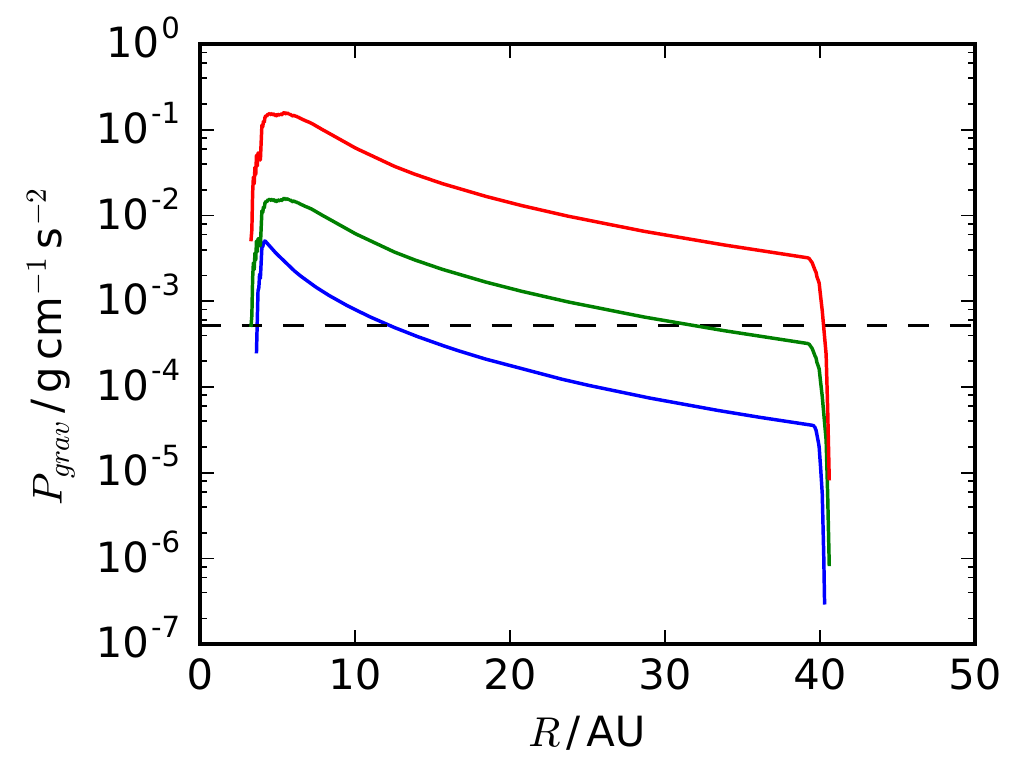}
	\caption[Gravitational pressure in the stellar disks as a function of radius.]{Gravitational pressure in the disk as a function of radius, as defined by Equation \ref{eqn5:gp}. The blue, green and red lines correspond to the low, medium and high values of $\rho_{0}$ (see Table \ref{table5:sims} for details). The dashed horizontal line shows the peak ram pressure of the SNR at $0.3\,$pc.}
	\label{fig5:radial_gp}
\end{figure}

\begin{figure}
	\centering
	\includegraphics[width=0.5\textwidth]{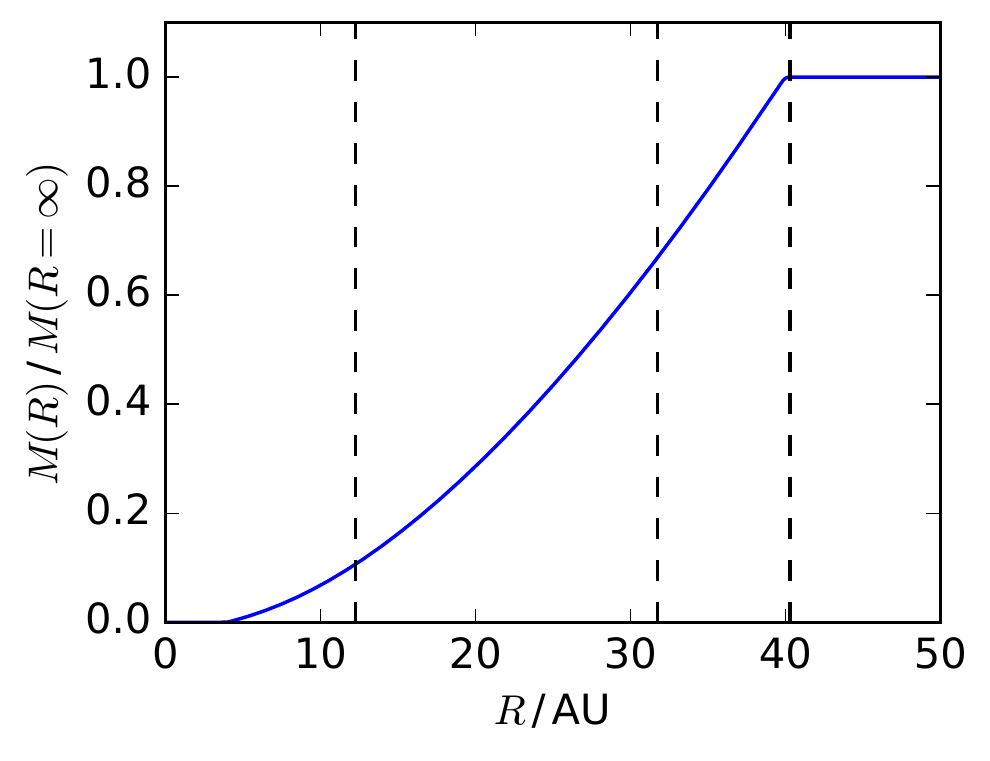}
	\caption[The fractional integrated mass of the disk as a function of disk radius.]{The fractional integrated mass of the disk as a function of disk radius. Note this is the same for all values of $\rho_{0}$ for the density distribution given by Equation \ref{eqn5:rhofull}. The dashed vertical lines shows the radius outside which the gravitational pressure of the disk is less than the peak ram pressure of the SNR for each value of $\rho_{0}$. From left to right the dashed lines are shown for low, medium and high disk mass cases.}
	\label{fig5:radial_mass}
\end{figure}

Material in the disk will be disrupted if the ram pressure of the wind exceeds the gravitational force per unit area, $P_{grav}$. This can be estimated as \citep{2000ApJ...538L.151C}:
\begin{equation}
	P_{grav} = \frac{GM_*\sigma}{R^2},
	\label{eqn5:gp}
\end{equation}
where $\sigma$ is the surface density of the disk. Figure \ref{fig5:radial_gp} shows the gravitational pressure as a function of radius for the three disk masses considered. By integrating surface density from the centre of the disk to the point at which the gravitational pressure drops below the peak ram pressure of the SNR, an estimate for the extent of the instantaneous stripping can be obtained.

Figure \ref{fig5:radial_mass} shows the cumulative integrated mass for the disk. If all the material at a gravitational pressure less than the peak ram pressure of the supernova is stripped, then the low mass disk is left with $10.7\%$ of its initial mass, the medium density disk retains $65.1\%$ and the high mass disk retains $99.95\%$.

In their investigation of the effects of inclination on the ablation of disk galaxies, \citet{2006MNRAS.369..567R} provide an argument for why stripping should be independent of inclination angle for small angles. Assuming the gas disk to be infinitely thin, the force due to ram pressure on a surface element $dA$ is $\rho_{wind} \nu_{wind}^{2} \mathrm{cos}(i)$. The ram pressure is effectively reduced by a factor of $\mathrm{cos}(i)$. As the radial gravitational force is balanced with the centrifugal force only the force perpendicular to the plane of the disk contributes. This is also reduced by a factor of $\mathrm{cos}(i)$. This means that the criteria for material to be stripped from the disk is independent of inclination angle. For highly inclined disks, the assumption of an infinitely thin disk will break down. While galactic disks are in a very different area of parameter space to stellar disks, these arguments are equally valid for either case.

%__________________________________________________________________

\section{Results}

\begin{figure*}
	\includegraphics[width=0.9\textwidth]{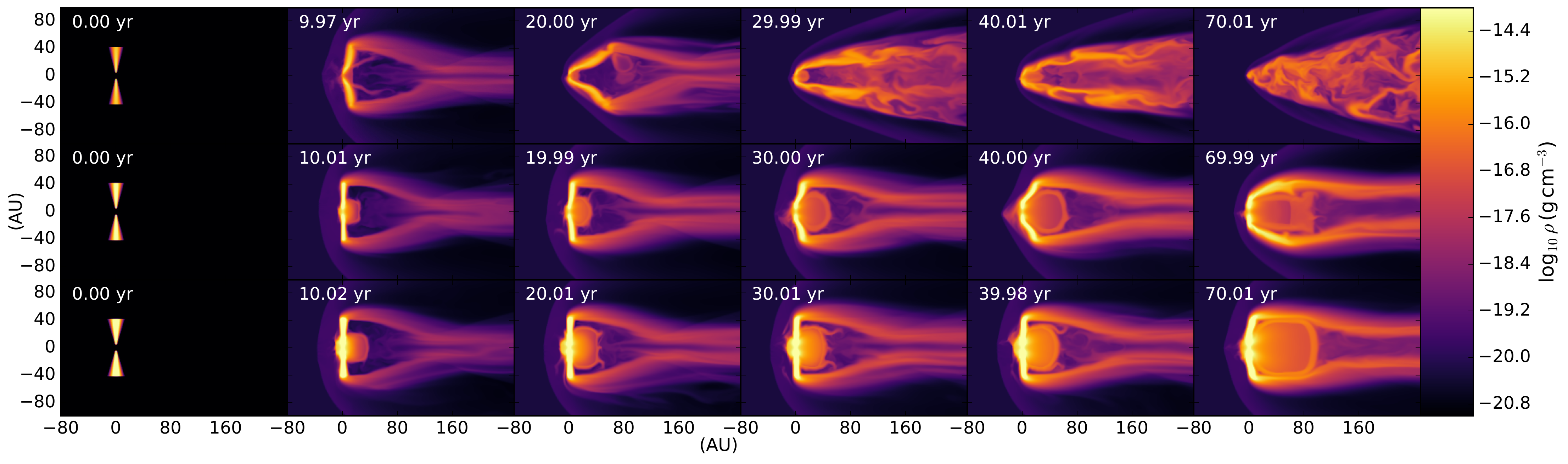}
	\caption[Slices through the X-Z plane at Y$=0$ for the simulations of varying disk mass.]{Slices through the X-Z plane at Y$=0$ for the simulations of varying disk mass. The simulations from top to bottom are \textbf{const00low}, \textbf{const00med} and \textbf{const00high}. The flow is constant and from left to right.}
	\label{fig5:rhosi_const_mass}
\end{figure*}

\begin{figure*}
	\includegraphics[width=0.9\textwidth]{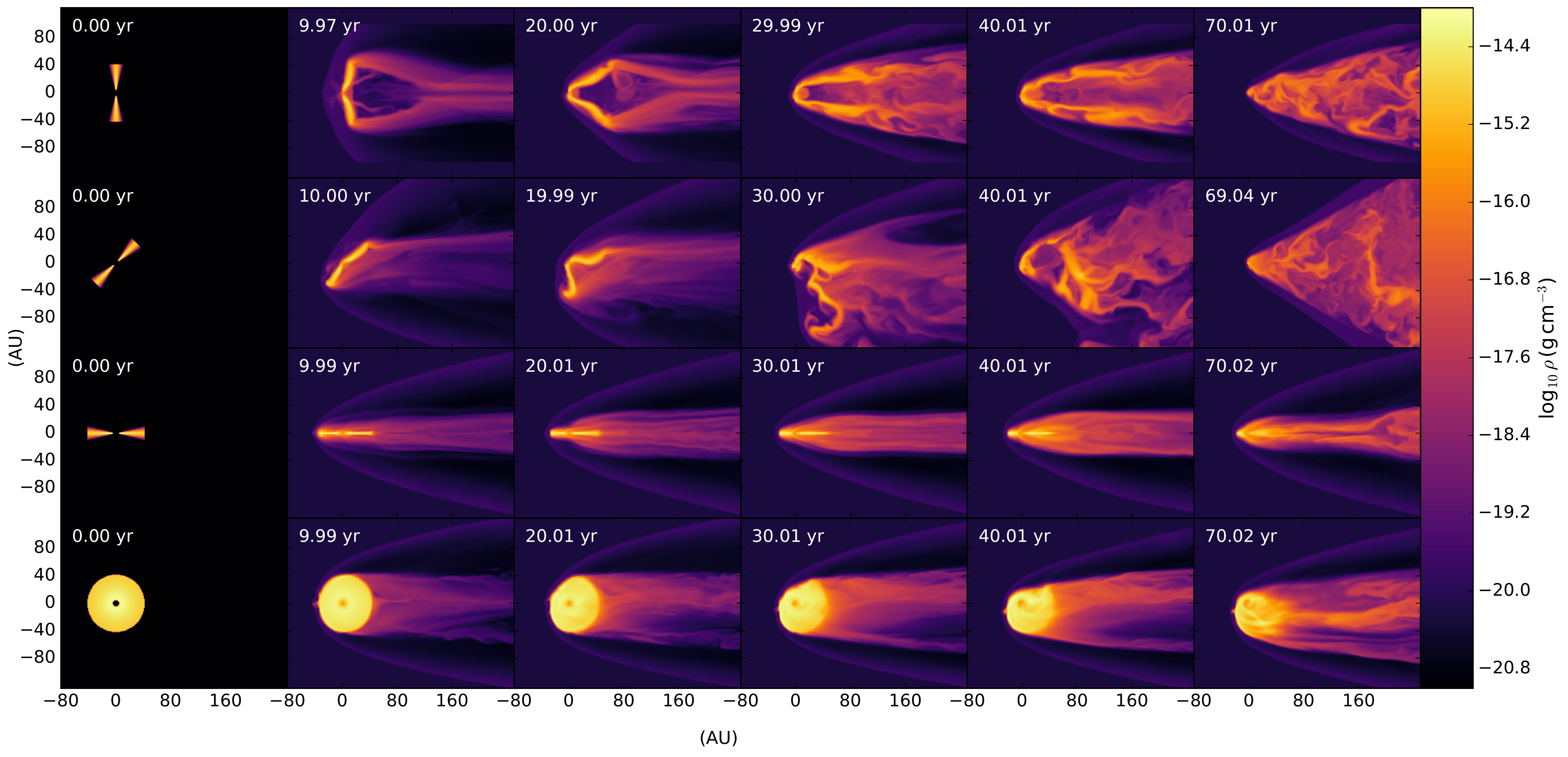}
	\caption[Slices through the X-Z plane at Y$=0$ for the constant flow simulations.]{Slices through the X-Z plane at Y$=0$ for the constant flow simulations. From top to bottom: \textbf{const00low}, \textbf{const45low} and \textbf{const90low}. The bottom row also shows a slice though the X-Y plane at Z$= 0$ for \textbf{const90low} (the disk rotates clockwise in these images). Note that although the simulations are performed by changing the angle of the flow, the images here are rotated such that the flow in each image is from left to right in the plane of the slice. The asymmetry in the bottom row arises due to the rotation of the disk.}
	\label{fig5:rhosi_const}
\end{figure*}

\begin{figure*}
	\includegraphics[width=0.9\textwidth]{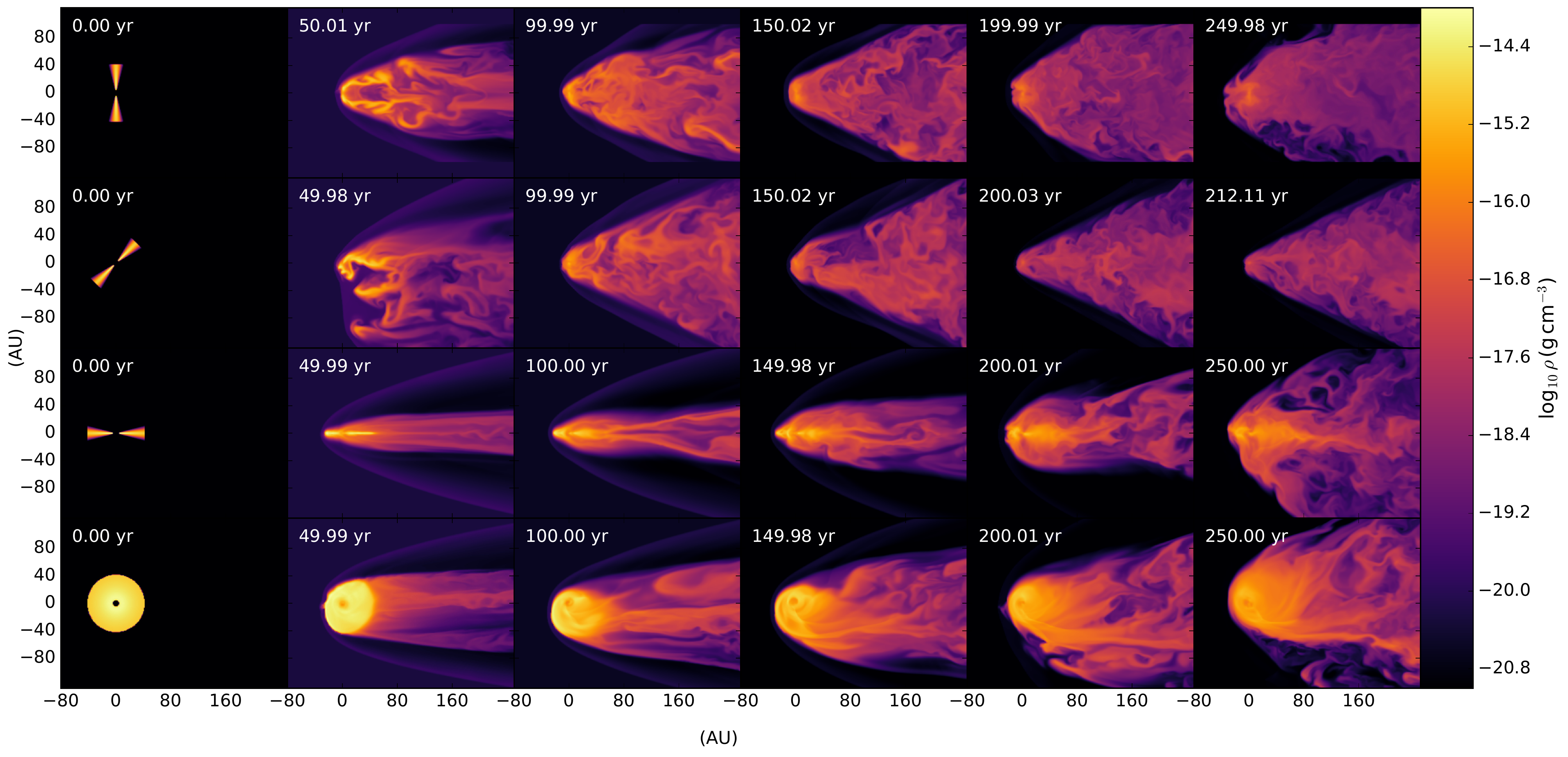}
	\caption[Slices through the X-Z plane at Y$=0$ for the dynamic flow simulations.]{As Figure \ref{fig5:rhosi_const} but for the dynamical flow simulations, \textbf{dyn00low}, \textbf{dyn45low} and \textbf{dyn90low}.}
	\label{fig5:rhosi_dyn}
\end{figure*}

Figures \ref{fig5:rhosi_const_mass}-\ref{fig5:rhosi_dyn} show snapshots of the evolution of the disk with two dimensional slices though the three dimensional grid. Figure \ref{fig5:rhosi_const_mass} shows the evolution of disks of different masses subject to a constant flow. As expected the more massive disks are more resilient to stripping. The low and medium mass disks are significantly deformed by the flow. For the low mass disk this breaks up the disk and forms a turbulent tail. However, in the medium mass case the deformed disk remains bound.

Figures \ref{fig5:rhosi_const} and \ref{fig5:rhosi_dyn} show the effect of inclination angle for the constant flow and dynamic flow cases respectively. For inclined disks the stripping is asymmetrical. For disk inclinations of $45^{\circ}$ the leading edge fragments and strips from the disk first, as the bow shock partially shields the trailing edge of the disk. As the disk continues to evolve the trailing edge is stripped and the disk becomes more symmetrical again. As the $45^{\circ}$ disk evolves, the disk is heavily stripped and eventually destroyed. At an inclination of $90^{\circ}$ the side of the disk rotating in the direction of the flow is stripped more heavily at first than the side rotating against the flow. Mass is stripped from the $90^{\circ}$ disk much more slowly than for disks with lower inclinations, due mainly to the lower cross-sectional area of the disk to the flow.

The morphology of the constant wind and dynamic wind cases is broadly similar in all cases. The main distinguishing feature is that as the pressure of the dynamic wind starts to decrease, the tail becomes much wider.

\begin{figure}
	\centering
	\includegraphics[width=0.5\textwidth]{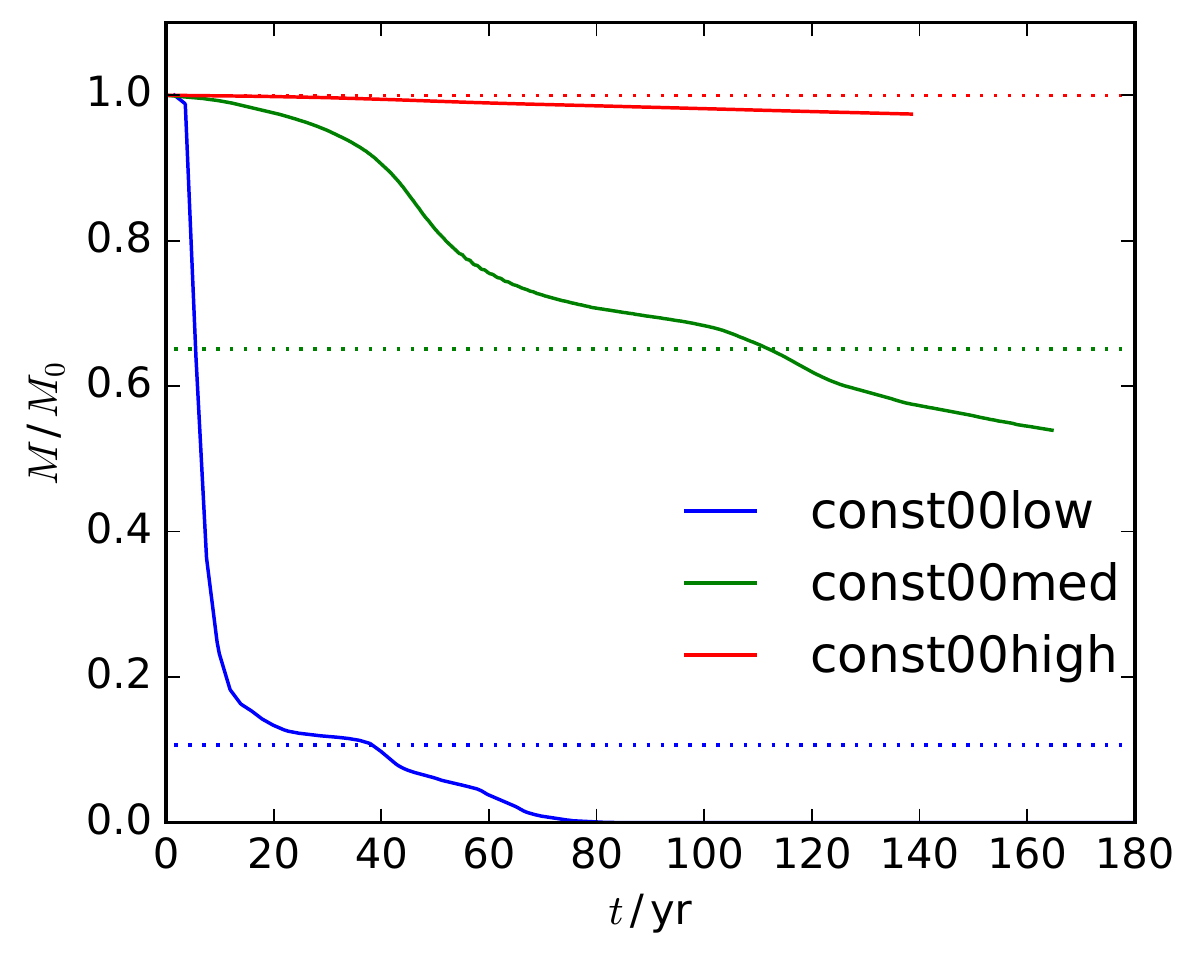} \\
	\includegraphics[width=0.5\textwidth]{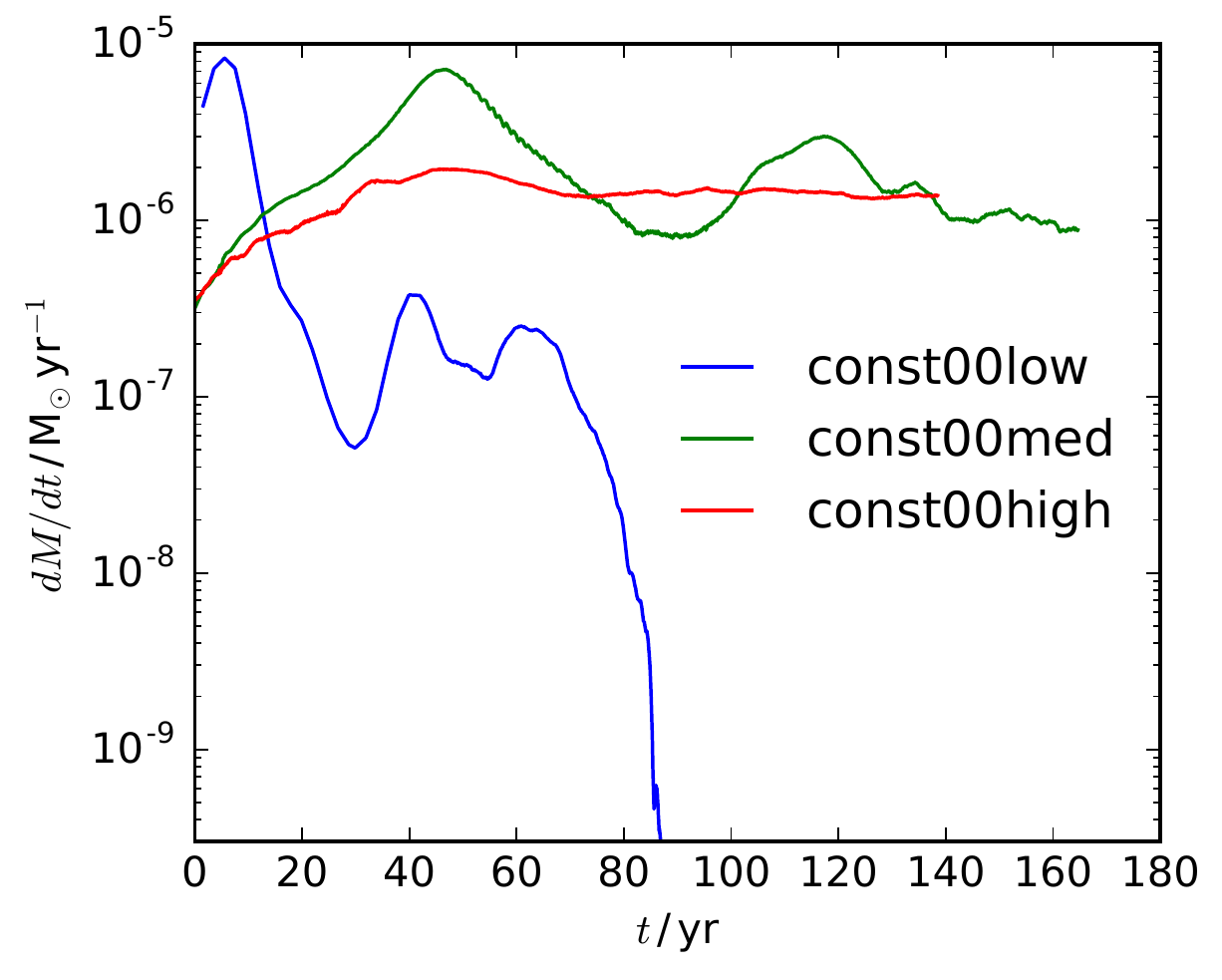}
	\caption[Mass of the disk as a function of time, for the three different disk masses simulated.]{The upper panel shows the mass of the disk as a function of time for constant face-on flows, showing the three different disk masses simulated. The dashed lines show the corresponding analytical predictions from Section \ref{sec5:analytic}. The lower panel shows the mass-loss rates for the same three simulations.}
	\label{fig5:massmass}
\end{figure}

\begin{figure}
	\centering
	\includegraphics[width=0.5\textwidth]{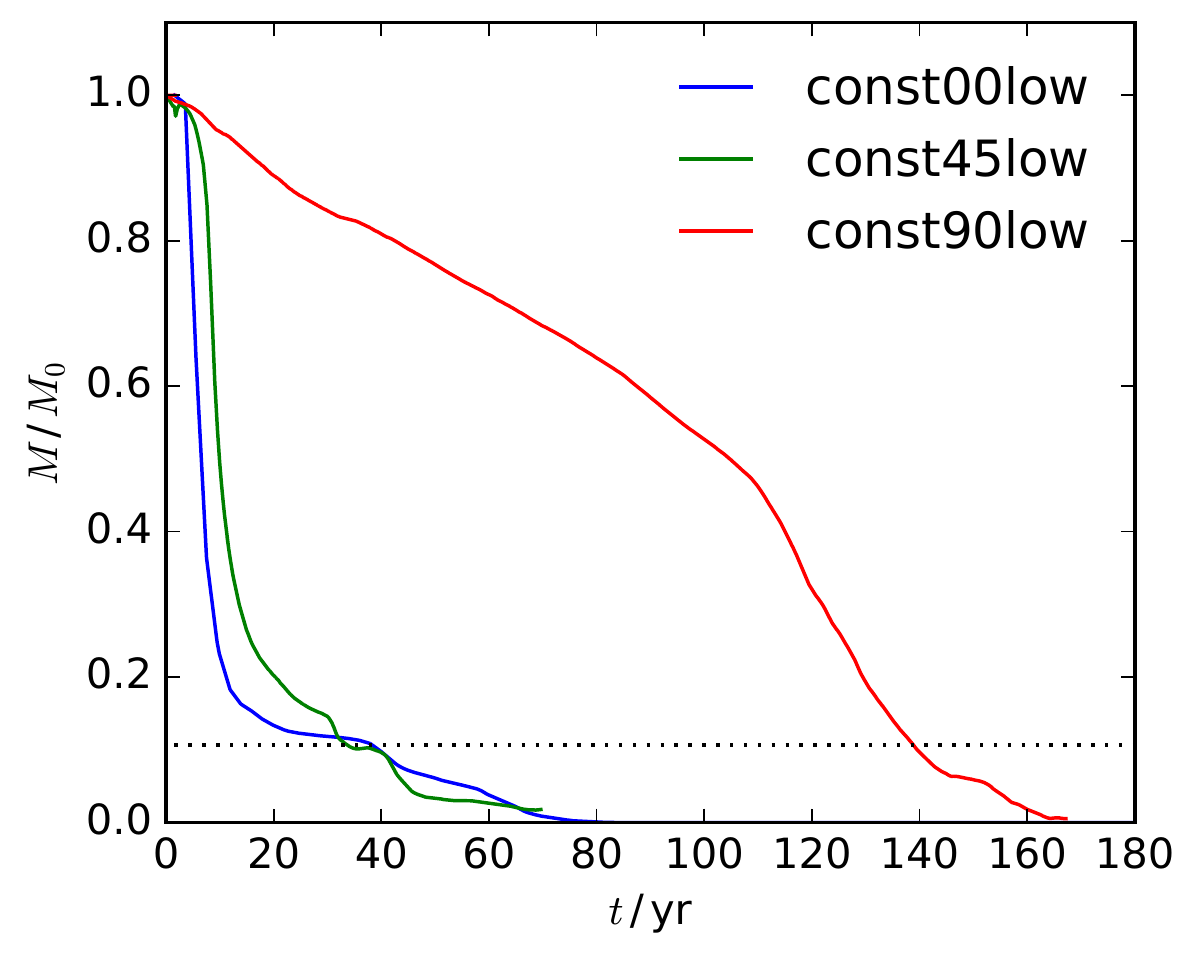} \\
	\includegraphics[width=0.5\textwidth]{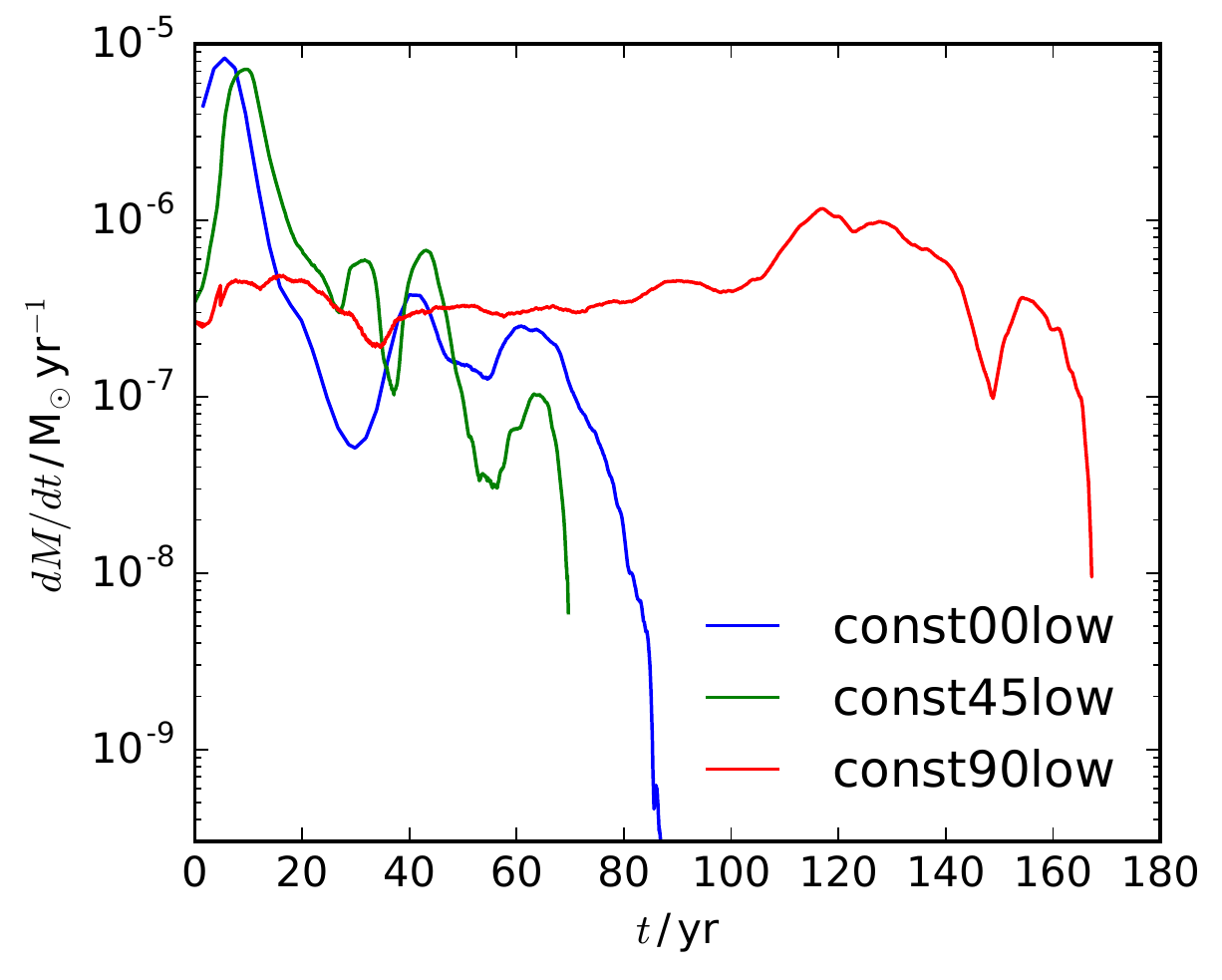}
	\caption[Mass of the disk as a function of time, for constant flows.]{As Figure \ref{fig5:massmass} but for the different inclinations of the low mass disk in a constant flow. The black dotted line shows the analytical predictions from Section \ref{sec5:analytic}.}
	\label{fig5:massconst}
\end{figure}

\begin{figure}
	\centering
	\includegraphics[width=0.5\textwidth]{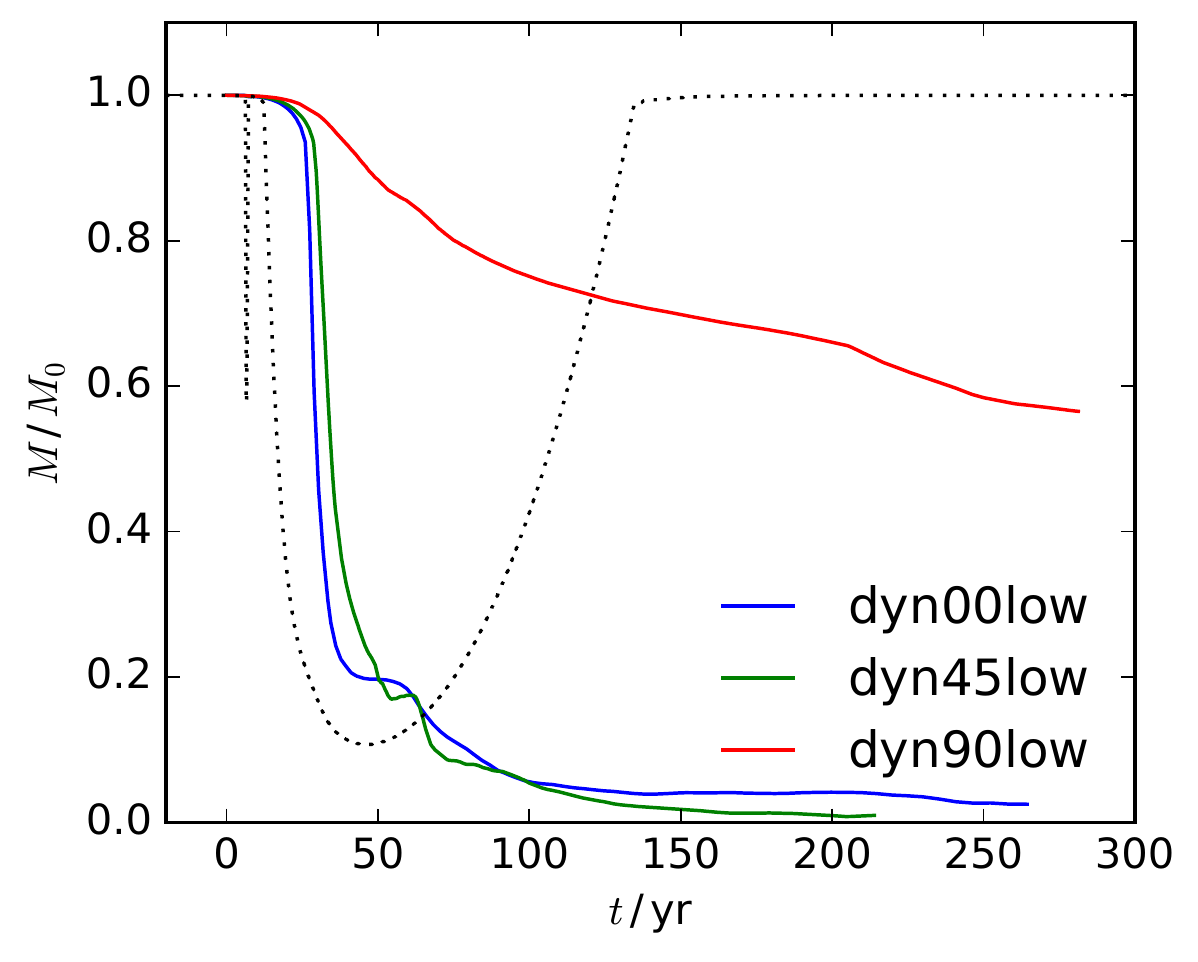} \\
	\includegraphics[width=0.5\textwidth]{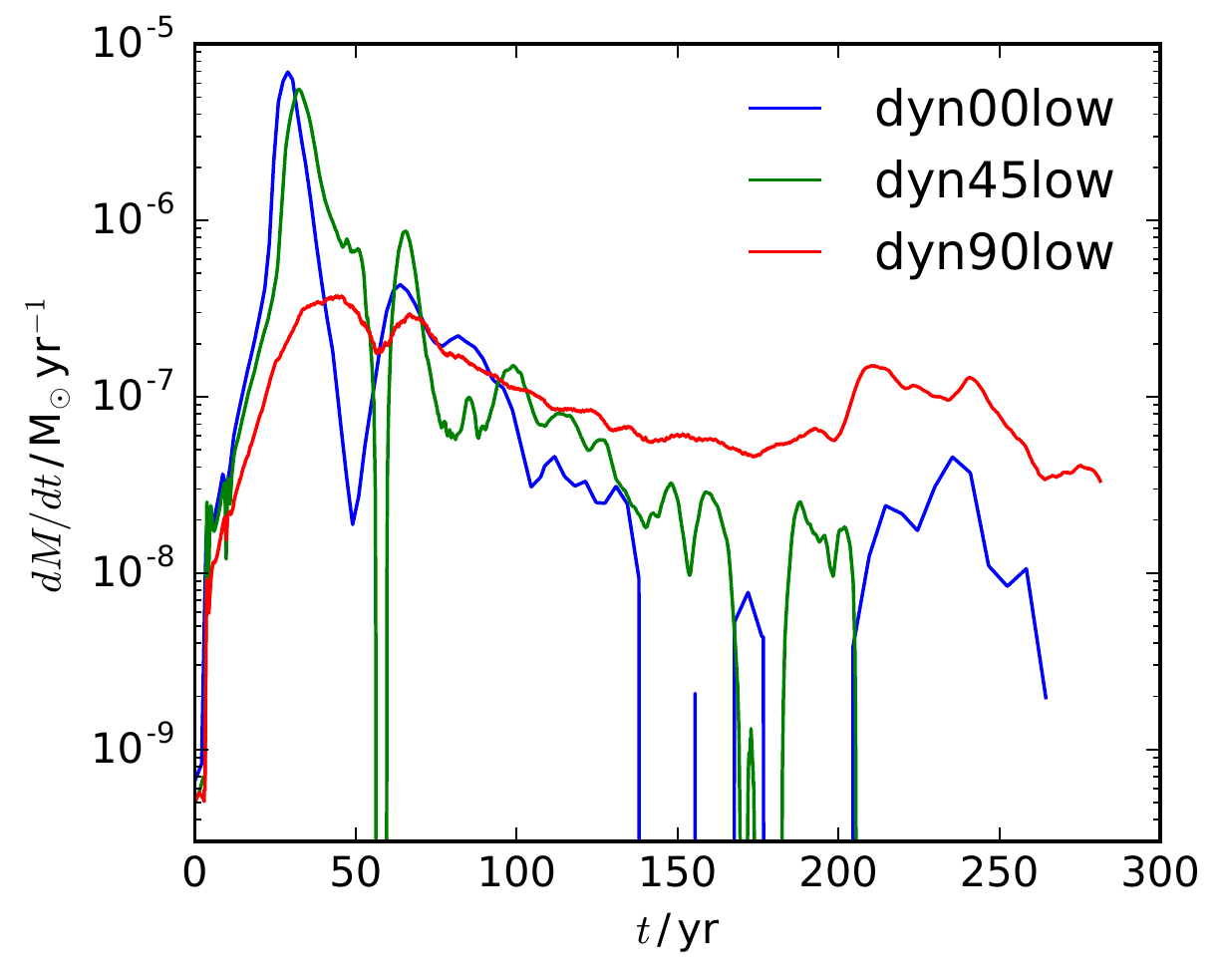}
	\caption[Mass of the disk as a function of time, for dynamic flows.]{As Figure \ref{fig5:massconst} but for dynamic supernova flow. The dashed black line represents the fraction of the initial disk mass that is vulnerable to stripping as the ram pressure evolves.}
	\label{fig5:massdyn}
\end{figure}

The amount of mass bound to the star's gravitational well is shown as a function of time in Figures \ref{fig5:massmass}-\ref{fig5:massdyn}. Simulations with a constant flow exhibit an initial rapid (``instantaneous'') stripping down to the level where the ram pressure is balanced by the gravitational pressure (where the disk mass plateaus), followed by a slower continual stripping caused primarily by the Kelvin-Helmholtz instability. Figure \ref{fig5:massmass} shows this for the three different disk masses simulated, together with the analytical approximations. For the low mass disk, the analytical prediction fits the position of the plateau very well. The medium mass disk begins to plateau but the continual stripping starts to dominate before it does so, pulling the mass below the analytical approximation. The high mass disk loses \correction{virtuallly} no mass due to instantaneous stripping and effectively starts in the continuous stripping phase. At the end of the simulation both the medium and high mass disks are losing mass at a rate of $\sim 10^{-6}\,$M$_{\sun}\,$yr$^{-1}$. During continuous stripping (between $t = 40\,$yr and $t = 70\,$yr), the mass-loss rate of the low mass disk is $\sim 10^{-7}\,$M$_{\sun}\,$yr$^{-1}$. While stripping is still occurring at the end of simulation, the constant ram pressure simulations are not continued beyond this point as the longer the simulation runs, the more unrealistic the assumption of peak ram pressure is. Running the simulations for $\approx 150-250$yrs provides enough data to establish trends and for comparison to other works.

As the inclination angle of the disk is increased, the stripping is generally slower (Figures \ref{fig5:massconst} and \ref{fig5:massdyn}). However, the difference between inclination angles of $0^{\circ}$ and $45^{\circ}$ is relatively small. Not only do they both plateau at about the same level, but they do so at about the same time (see Figures \ref{fig5:massconst} and \ref{fig5:massdyn}). This is because the flow in this instance is strong enough to deform the disk such that the initial inclination angle is no longer relevant. At high inclination angles a different behaviour is observed, and the disk survives significantly longer at an inclination of $90^{\circ}$. The same general trend can be seen in Figure \ref{fig5:massdyn} for the dynamic flow. Interestingly, for the face-on flow, the mass plateaus before the peak ram pressure has been reached, indicating that the history of the flow is important in shaping the disk and determining whether it is susceptible to ablation. The low mass disk only survives the dynamic flow when placed edge-on to the flow, retaining nearly $60\%$ of its mass in this case (Figure \ref{fig5:massdyn}). The edge on disk is losing $\sim 2 \times 10^{-7}\,$M$_{\sun}\,$yr$^{-1}$ at the end of the simulation. If this rate were to be sustained the disk would be destroyed in $\sim 300$yr. However, the ram pressure decreases by two orders of magnitude in this time, so there will be some but not complete stripping during this period. The disk is therefore expected to survive significantly longer than we have simulated.

\begin{figure}
	\centering
	\includegraphics[width=0.5\textwidth]{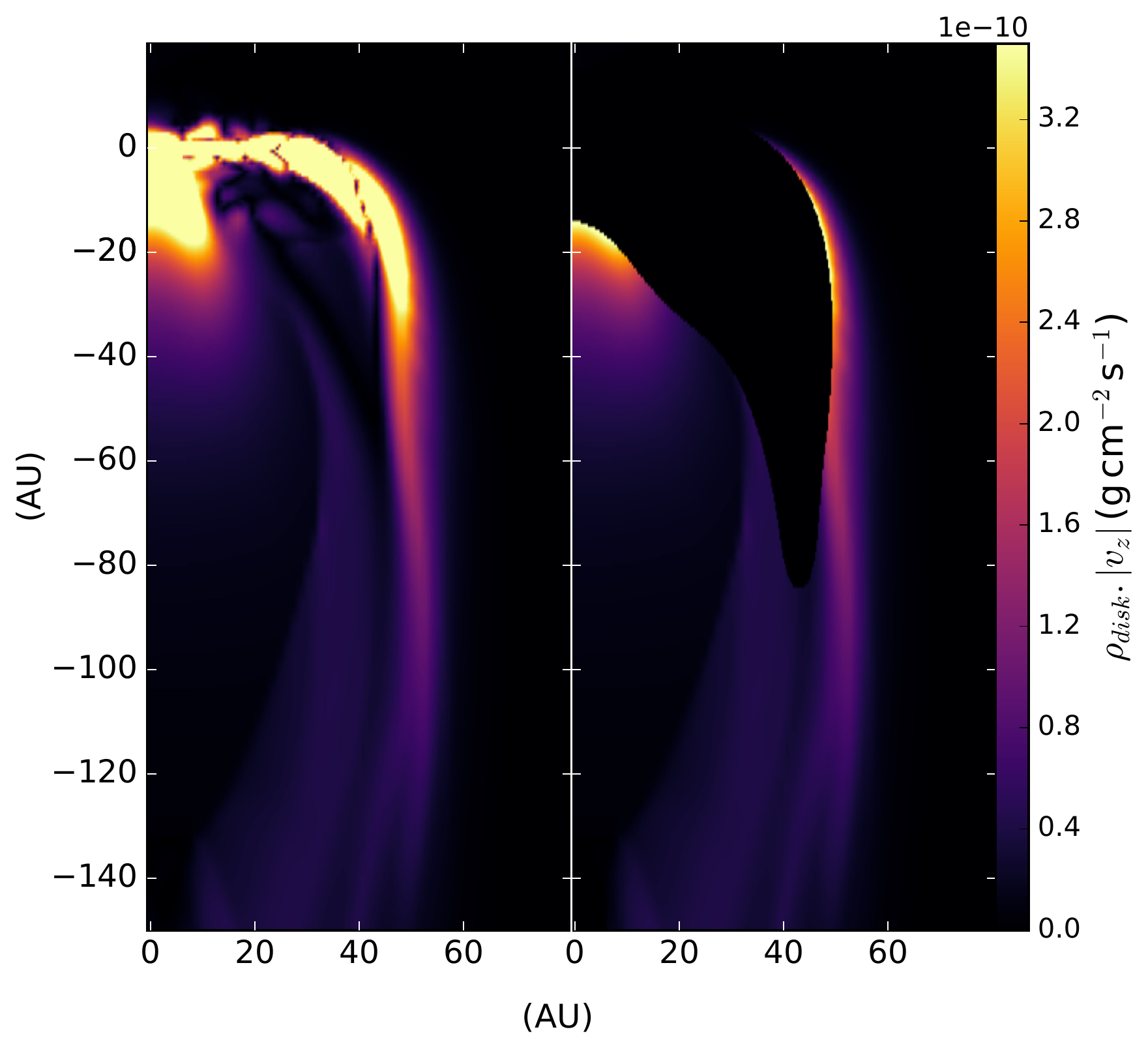}
	\caption[Mass flux in the X-Z plane]{Absolute value of the mass flux in the X-Z plane for \textbf{const00high} at $t = 140\,$yr. The left plane shows the total contribution to the mass flux from the disk material. The right plane shows only disk material which is gravitationally unbound from the central star.}
	\label{fig5:massflux}
\end{figure}

\begin{figure}
	\centering
	\includegraphics[width=0.5\textwidth]{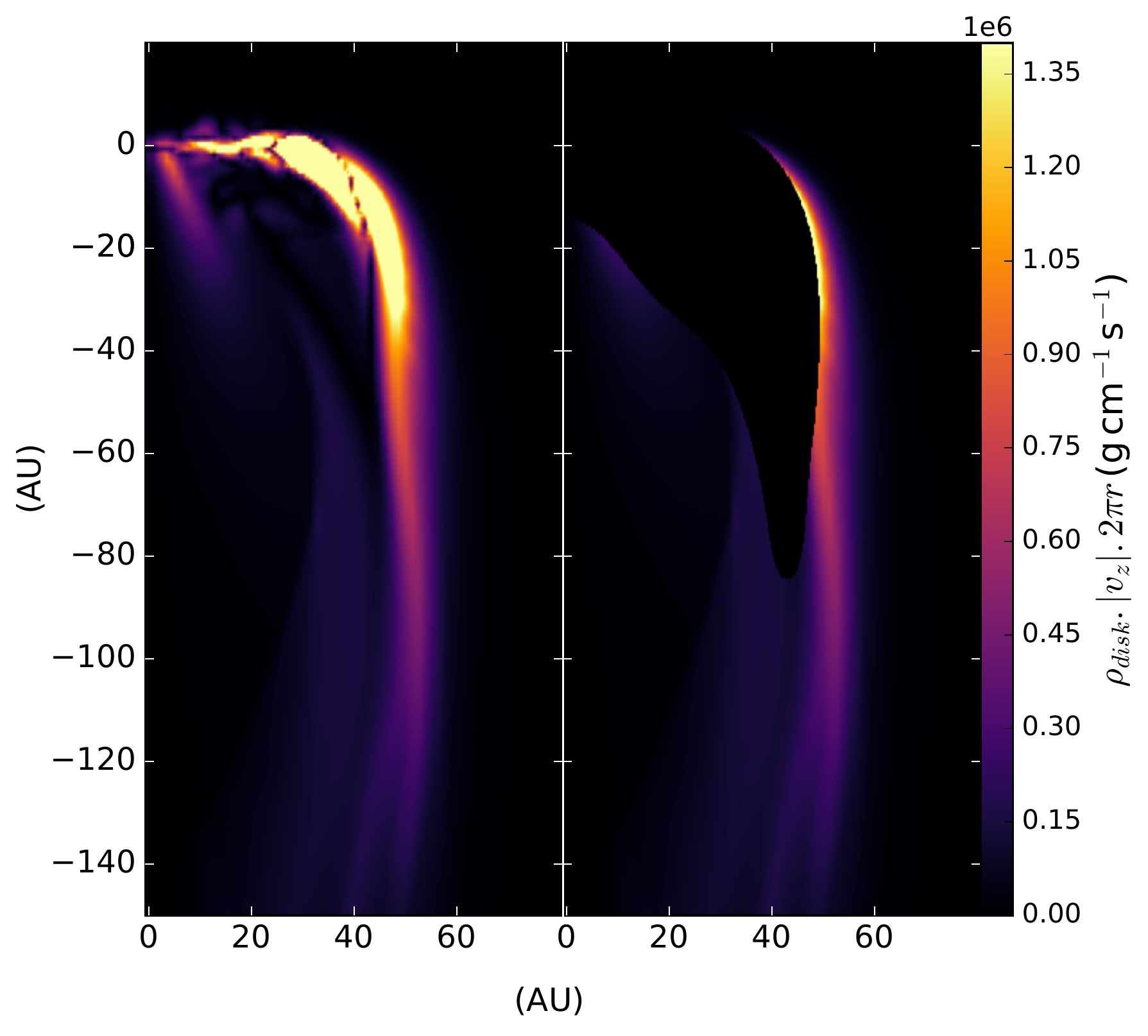}
	\caption[Mass flux in the X-Z plane]{As Figure \ref{fig5:massflux}, but showing the mass flux multiplied by a factor of $2 \pi r$.}
	\label{fig5:massflux_az}
\end{figure}

Figure \ref{fig5:massflux} shows the z-direction mass flux of the disk material at the end of simulation \textbf{const00high}. To show the regions where the disk is losing mass the unbound material is shown in the right panel. Mass is lost from two main regions: along the sides of the deformed disk and also though the central hole. While the mass flux in both these areas is similar the central region spans a much smaller area. Figure \ref{fig5:massflux_az} shows the same mass flux as Figure \ref{fig5:massflux} now multiplied by a factor of $2 \pi r$ to account for the differing contributions to the total mass flux. It is clear that the majority of the mass is lost from the edge of the disk. Ablation from the centre of the disk accounts for approximately $30\%$ of the mass being ablated at this point in the disk's evolution.

\begin{figure*}
	\centering
	\includegraphics[width=0.49\textwidth]{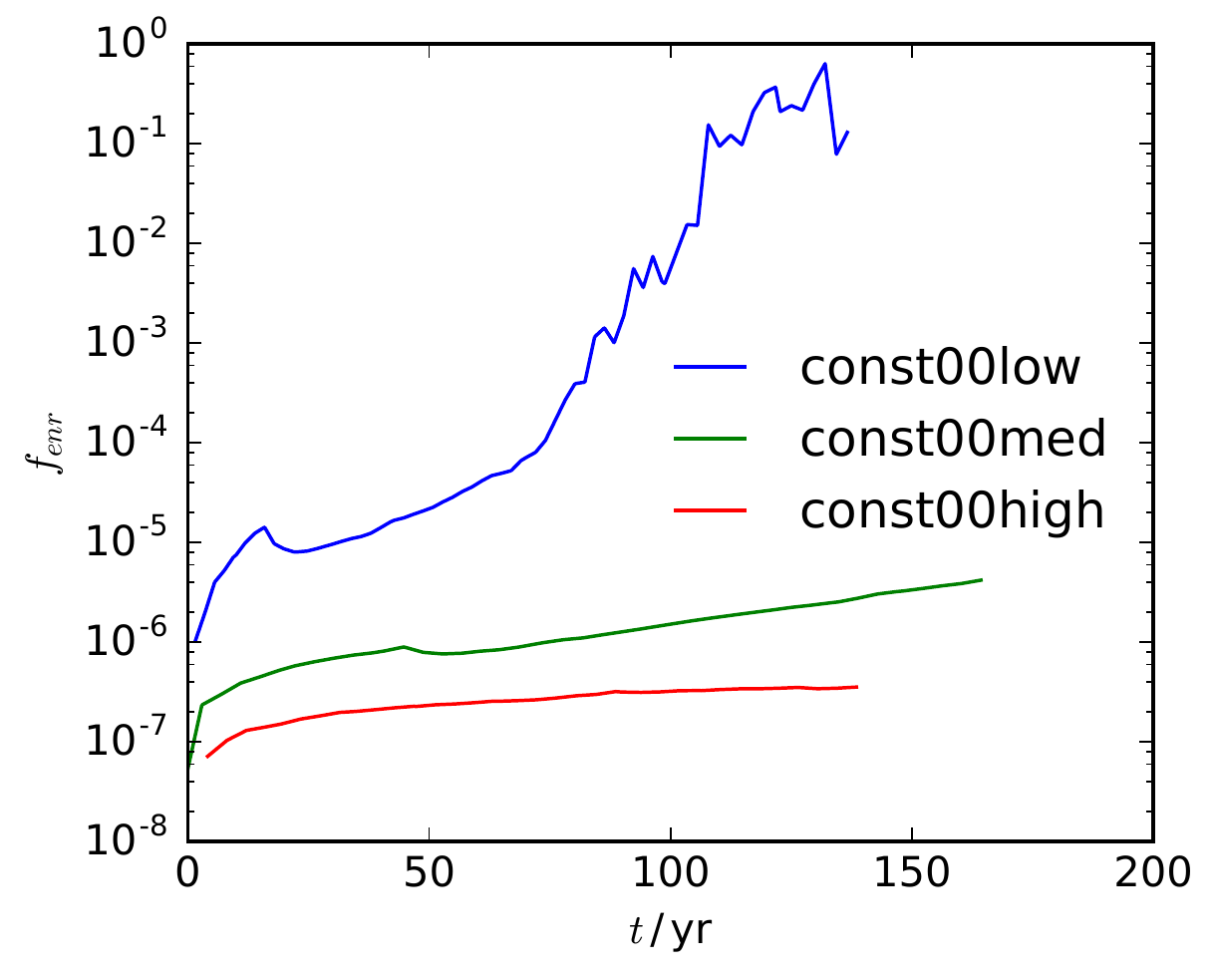}
	\includegraphics[width=0.49\textwidth]{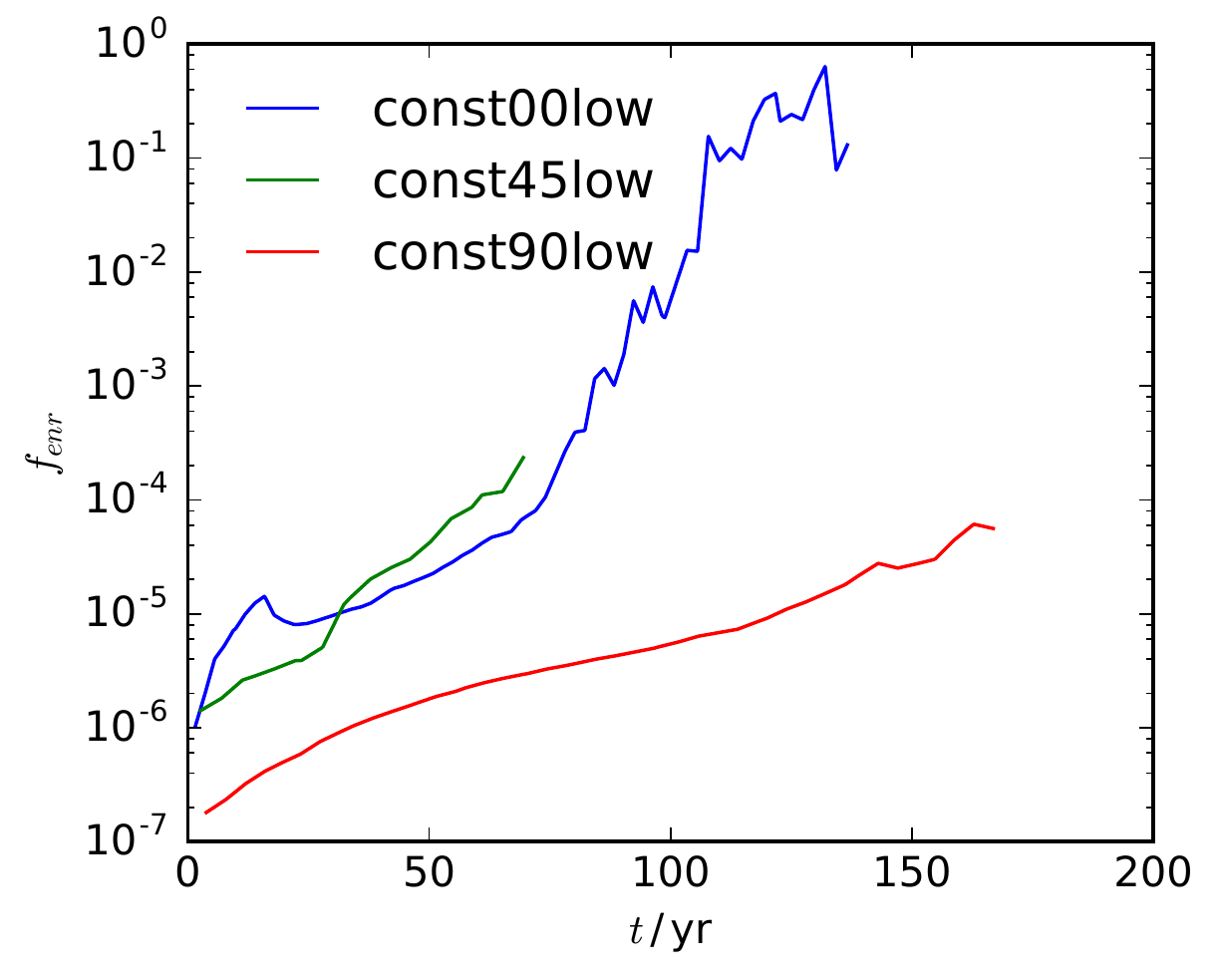} \\ 
	\includegraphics[width=0.5\textwidth]{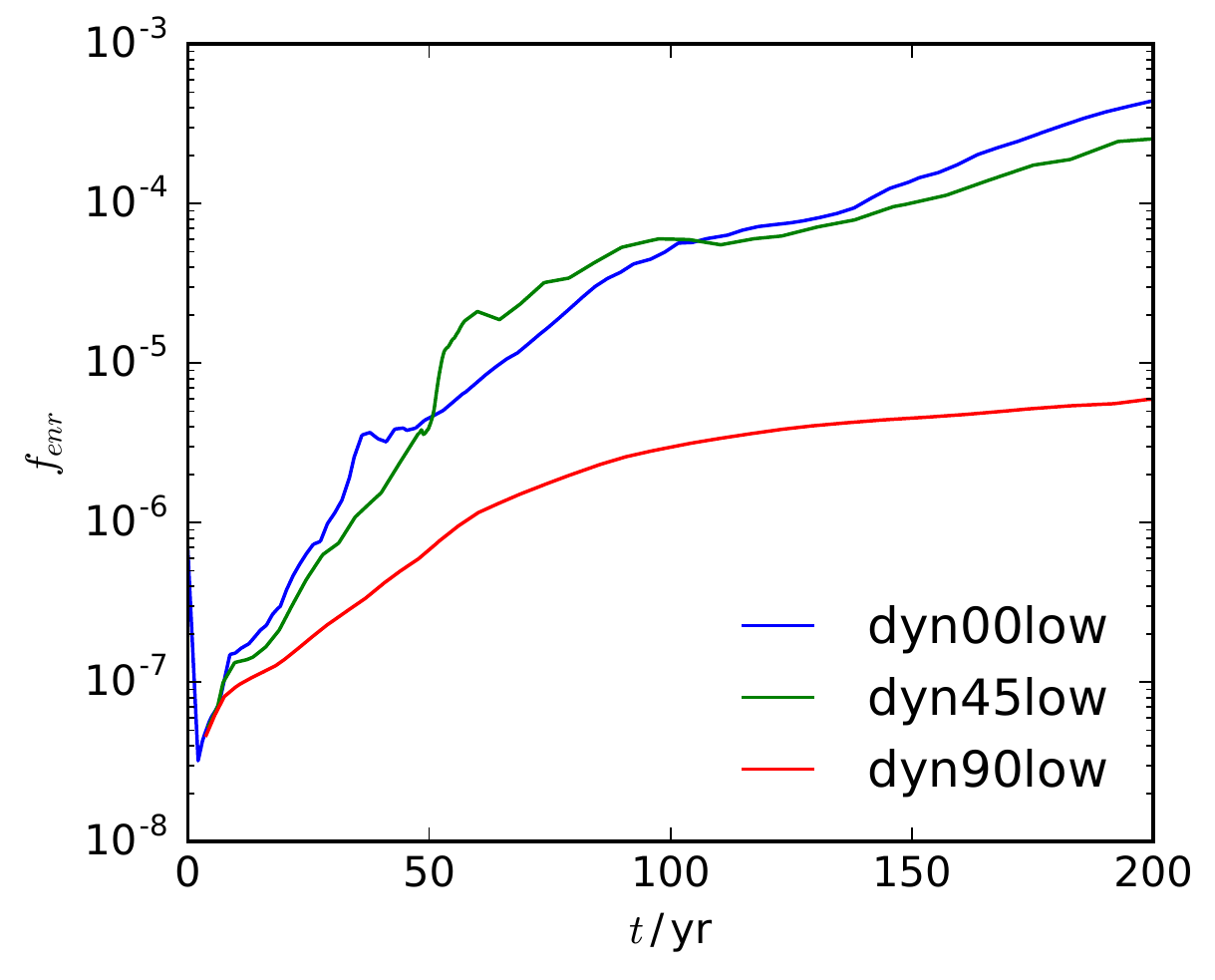}
	\caption[SNR mass fraction bound to disk]{\correction{Enrichment fraction ($f_{enr}$, see \ref{sec5:def}) as a function of time. The only disks to survive until the end of simulation are \textbf{const00high}, \textbf{const00med} and \textbf{dyn90low}. The relativly high enrichment fractions of the other disks (particularity \textbf{const00low}) at late times can be attributed to a dramatic reduction in $M_{disk}$.}}
	\label{fig5:snrbound}
\end{figure*}

Figure \ref{fig5:snrbound} shows the amount of SN ejecta that becomes bound to the disk, tracked using an advected scalar. The disks in simulations \textbf{const00med}, \textbf{const00high} and \textbf{dyn90low} are the only disks to survive to the end of simulation and the only ones that could be considered to be enriched by SN ejecta, although all simulations are shown for completeness. The mass of SN ejecta in the disk is $\sim10^{-9}\,$M$_{\sun}$ for the two face-on disks and $\sim10^{-10}\,$M$_{\sun}$ for the edge-on disk. 

\begin{figure*}
	\centering
	\includegraphics[width=0.49\textwidth]{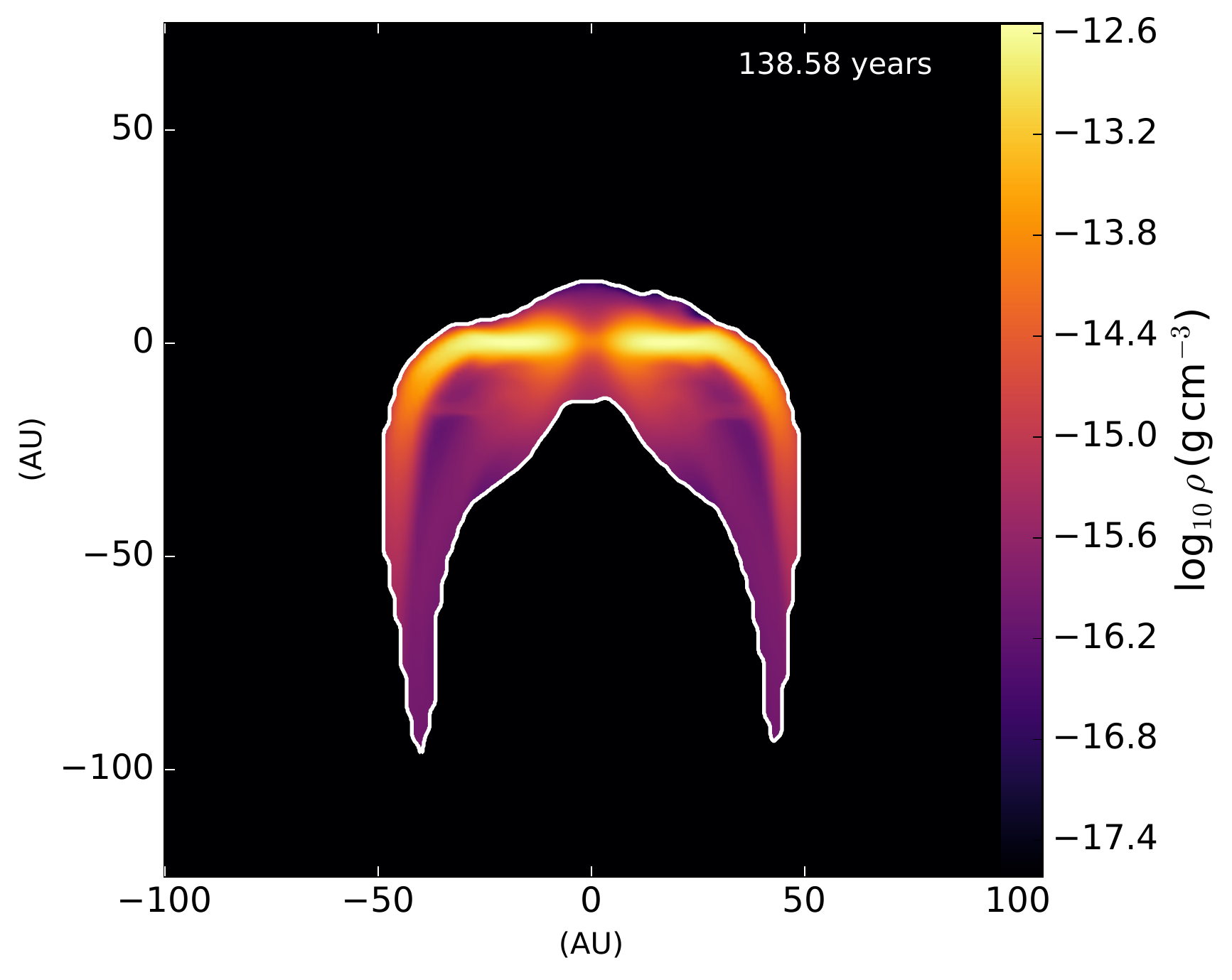}
	\includegraphics[width=0.49\textwidth]{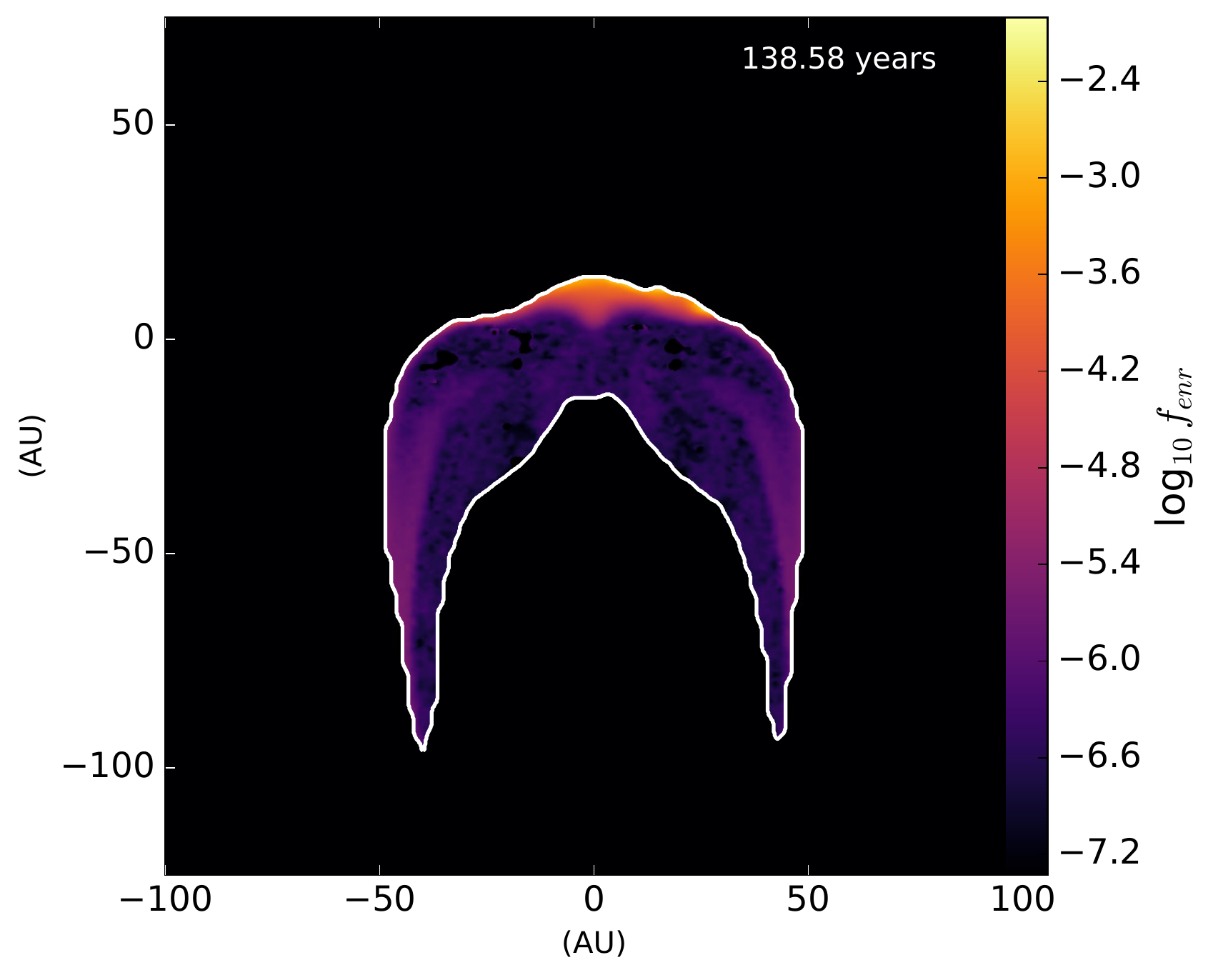}
	\caption[]{A slice though the X-Z plane at the end of simulation \textbf{const00high}, showing only material bound to the central star's gravitional potential. The left image shows the total mass density. The right image shows the value of the advected scalar used to trace supernova ejecta material. \correction{It is equivalent to the enrichment fraction on a per cell basis (see \ref{sec5:def}).} The edge of the bound region is highlighted in white to aid comparison between the two images.}
	\label{fig5:constmedbound}
\end{figure*}

\begin{figure*}
	\centering
	\includegraphics[width=0.49\textwidth]{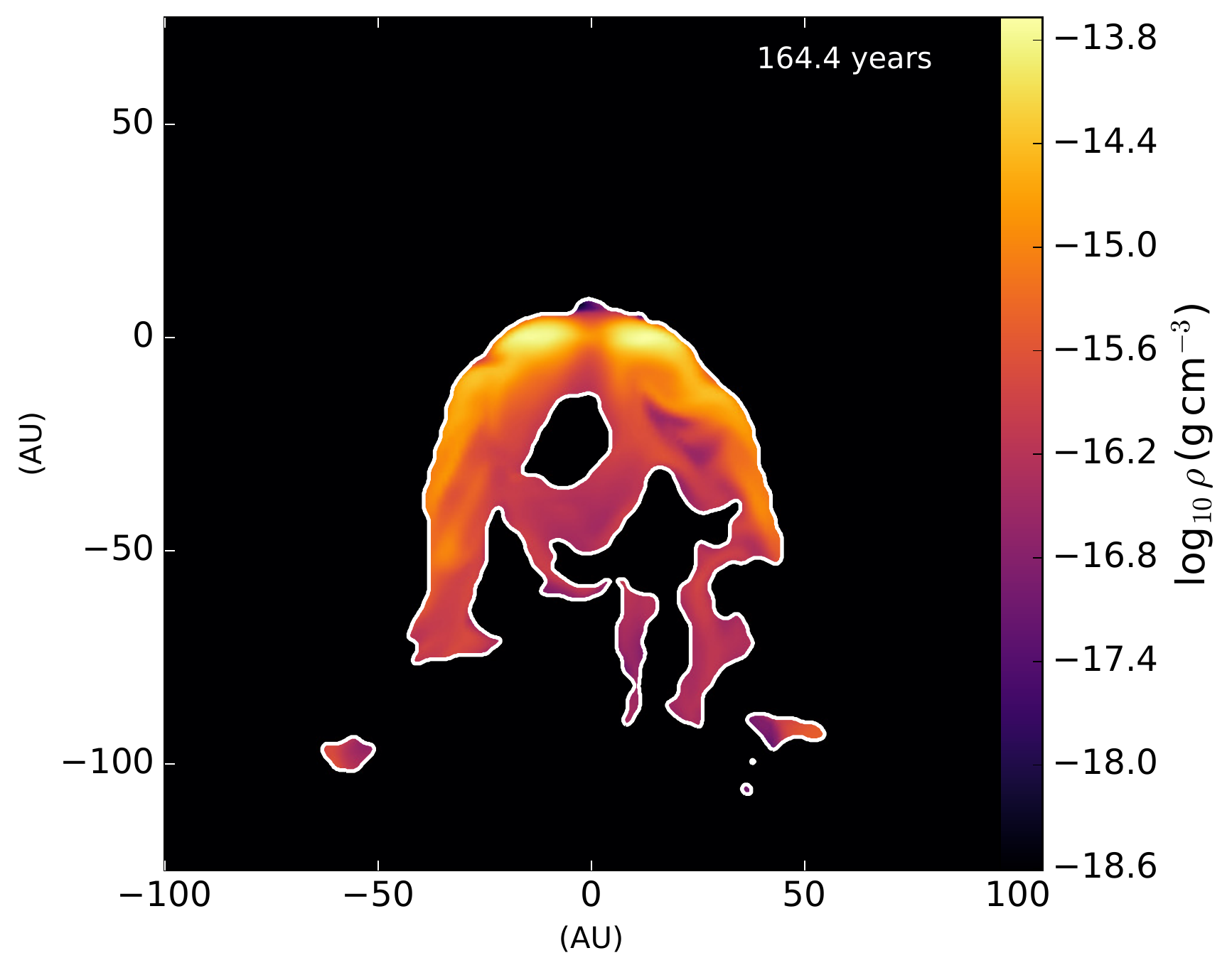}
	\includegraphics[width=0.49\textwidth]{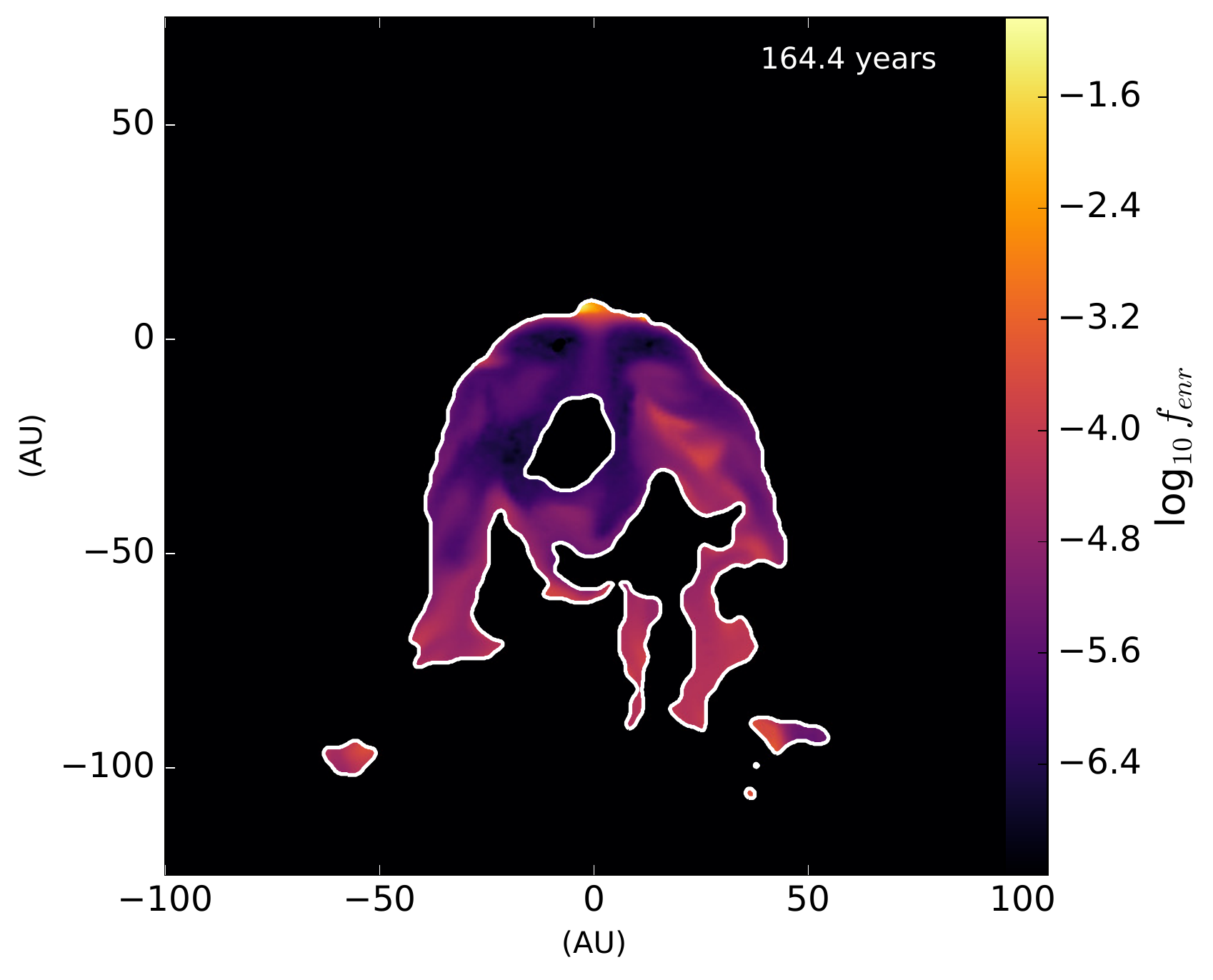}
	\caption[]{As Figure \ref{fig5:constmedbound} but for \textbf{const00med}.}
	\label{fig5:consthighbound}
\end{figure*}

\begin{figure*}
	\centering
	\includegraphics[width=0.49\textwidth]{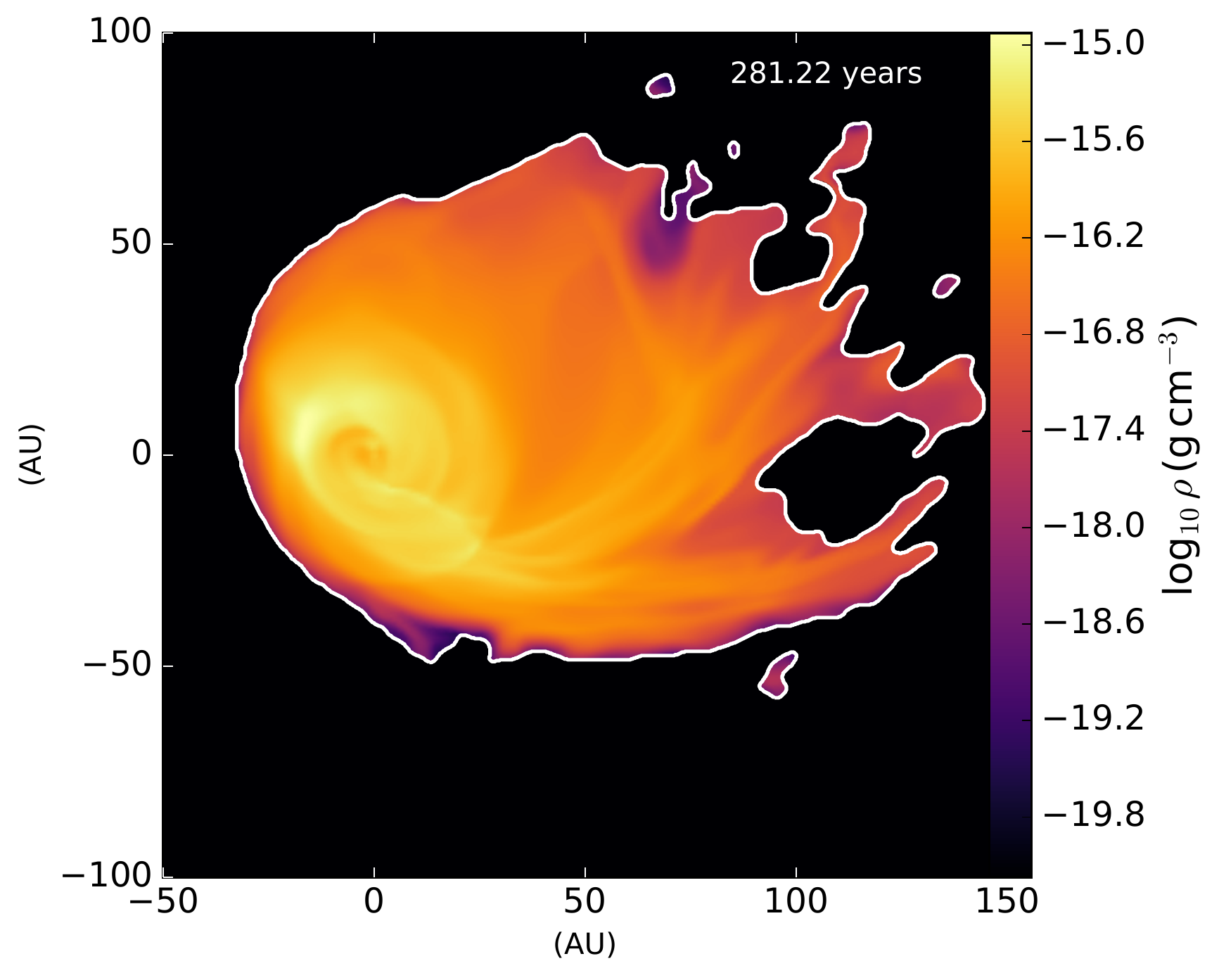}
	\includegraphics[width=0.49\textwidth]{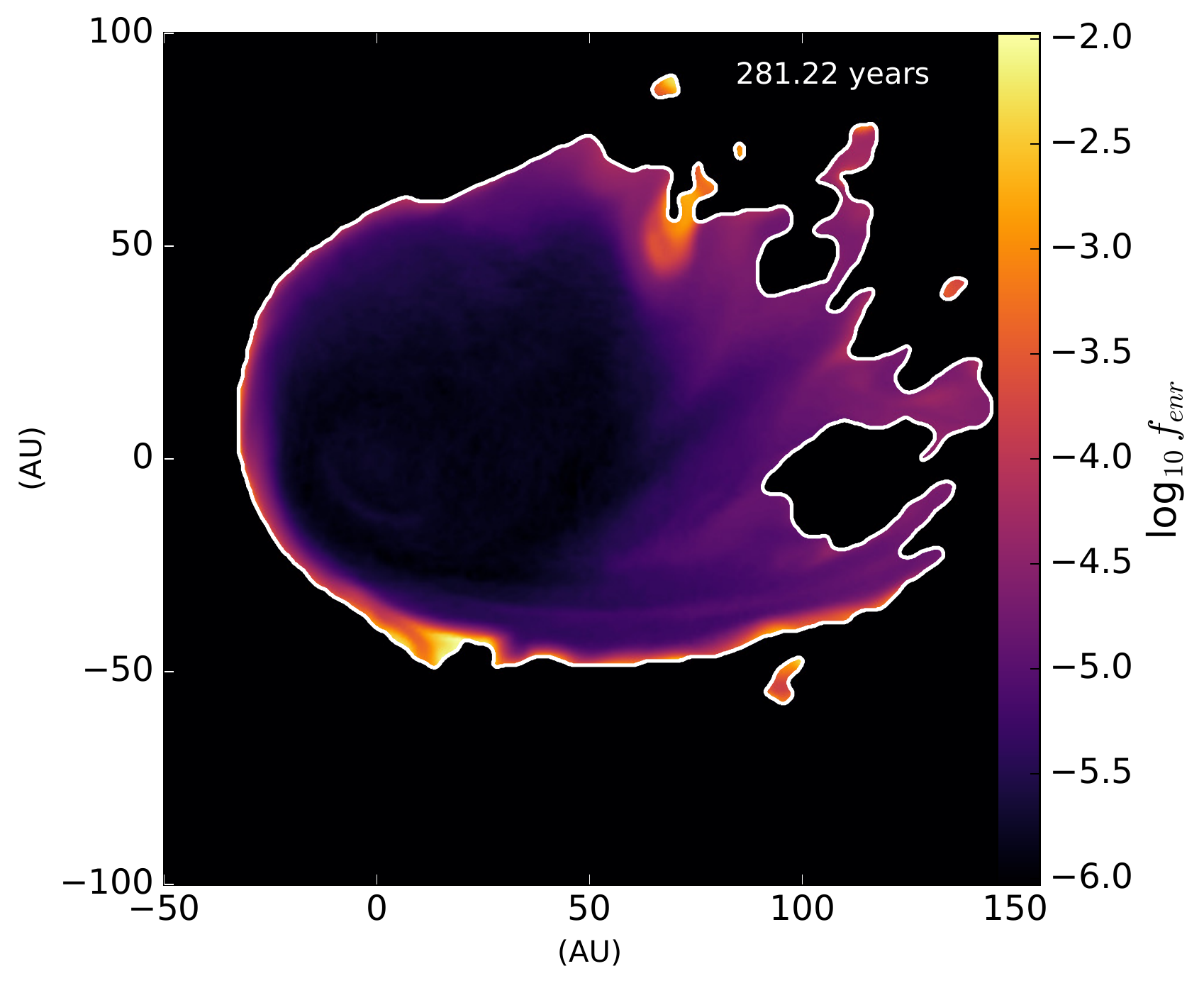}
	\caption[]{As Figure \ref{fig5:constmedbound} but for \textbf{dyn90low}, and showing slices though the X-Y plane.}
	\label{fig5:dyn90bound}
\end{figure*}

Figures \ref{fig5:constmedbound}-\ref{fig5:dyn90bound} show the distribution of bound material for the three disks that survive until the end of the simulation. The right panels in these figures show the value of the advected scalar used to trace supernova ejecta: a value of 1.0 indicates pure ejecta, and a value of 0.0 indicates that no supernova ejecta is present. In all 3 figures the bound supernova ejecta is predominantly found on the upstream face (or edge) of the disk.
For the edge-on disk the SN ejecta which becomes bound enters an orbit around the edge of the disk. The ejecta  concentration on the leading surface is of order 1 part in $10^{4}$ (in some small localised low-density regions the concentration is at the 1 per cent level). Inside the disk the concentration drops to typically 1 part in $10^5-10^6$. Thus we see some ejecta material being deposited onto the surface of the disk, but little mixing into the disk interior on the timescales of our simulations. While our simulations do not run long enough to capture the subsequent mixing of this material (nor are turbulent motions within the disk properly resolved), \cite{2013ApJ...773....5B} provide good estimates of the timescales over which mixing is likely to happen, finding that homogenization in a $40$AU disk occurs on timescales of $10^3$ to $10^4$yr. No significant decay of SLRs would be expected over this time frame.

%__________________________________________________________________

\section{Discussion}

\subsection{Comparison to previous works}

The main point of comparison is with \citet{2005ASPC..341..527O}. Our face-on high density disk can be considered equivalent to their canonical disk. Good agreement is obtained in that virtually no stripping is seen in either case. However they see less than $0.1\%$ mass lost over $2000\,$yr whereas we see $\sim2.5\%$ lost over $140\,$yr. This difference can be attributed to the fact that our high density disk is only simulated for the constant ram pressure flow where the long term ablation is much stronger, whereas \cite{2005ASPC..341..527O} used a time varying flow. Our simulations also differ in that we observe consistency with the analytical predictions of  \citet{2000ApJ...538L.151C}, whereas they state that the bow shock shields the disk causing it to be able to survive higher levels of ram pressure.

\citet{2014MNRAS.444.2884L} study the effect of triggered star formation, allowing a disk to form and evolve under the effects of a wind. Their simulations are in a very different region of parameter space to ours: the ram pressure of their wind is $\sim7$ orders of magnitude lower, so they are observing the interaction on a timescale of Myrs and their disk after formation has a radius of $\sim1000$au, with a resolution of $23$au. However, they do note that at the end of their simulation the disk radius is much less than that predicted using the method of \citet{2000ApJ...538L.151C}.

As there have been no studies specifically of the ablation of inclined stellar disks, the only comparison that can be made is with galactic disks. \citet{2006MNRAS.369..567R} and \citet{2009A&A...500..693J} both specifically investigated the effect of inclination on the ram pressure stripping of galactic disks. We observe the same general dependence on inclination angle, where stripping is similar for low inclination angles and is only strongly impeded for close to edge-on orientations. We also observe the same asymmetry when disks are not face on. \citet{2009A&A...500..693J} argues that the parallel side is stripped more easily as the wind works in the direction of rotation to push the material off the disk, whereas the material on the anti-parallel side must first be slowed to zero velocity and re-accelerated in the opposite direction if it is to be stripped on that side on the disk (otherwise it will continue to orbit round to the other side of the galaxy where it may then be stripped more easily). \citet{2006MNRAS.369..567R} note that after an outer rotation period the disc becomes symmetrical again, as the entire disk has experienced the side with stronger stripping.

For galactic disks the stripping occurs over a relatively long period of time, with the disk experiencing several rotation periods while being ablated. However, the nature of the stripping depends on how the galaxy is orbiting within the cluster. A galaxy on a circular orbit will experience constant ram pressure throughout its lifetime, whereas a galaxy falling radially will experience a shorter peak of ram pressure. \citet{2007MNRAS.380.1399R} calculate the dynamic ram pressure for several different orbital paths a galaxy might take. Defining a significant ram pressure as one that would be expected to strip at least half of the disk mass if maintained constantly (this corresponds to what \citet{2007MNRAS.380.1399R} refer to as a "medium" ram pressure and is $\sim10^{-11}\,$g$\,$cm$^{-1}\,$s$^{-2}$ in their case), the galactic disk experiences significant ram pressure for approximately $0.75$ to $2$ times the outer rotation period during each orbit. 

For the supernova and stellar disk interaction examined here the ram pressure is only significant for $\sim1/3$ of the outer rotation period. This means it is possible for the stellar disk to maintain asymmetry after the point where stripping is significant, although it is not clear where the boundary lies, either in terms of ram pressure or rotation periods, in order for this to occur. Also note that this effect requires the flow to impact the disk edge-on or near edge-on for extended periods, whereas a galaxy is likely to experience stripping from a number of different angles at different points in its orbit.

\subsection{Internal stripping}

Our simulations show that a significant amount of mass is lost though the central hole of the disk during the continuous stripping phase. Theoretical considerations show that large planets can cause gaps to form in the disk \citep{1996ApJ...460..832T} and observations of protoplanetary disks have since detected the presence of such gaps \citep{2002ApJ...568.1008C, 2016ApJ...820L..40A}. If these gaps are large and deep enough they could provide additional channels for ablation to take place.

Similarly, in disk galaxy simulations cooling causes the disk to fragment and allows the wind to flow though areas of low density within the galaxy \citep{2009ApJ...694..789T}. This causes the instantaneous stripping to proceed faster but to the same radial extent compared to the adiabatic, non-fragmented disk. The differences in long term stripping were not investigated but it is likely to be accelerated in the same way by the additional surface area, and will almost certainly effect the morphology of the flow. However, cooling is not significant in the bowshock surrounding the disk in our simulations. This may change if the SNR shock itself became strongly cooling (due to a greater age at the time of interaction, and/or higher circumestellar densities).

\subsection{Continuous stripping}

A number of different authors \citep{1982MNRAS.198.1007N, 1986MNRAS.221..715H, 1997ApJ...484..810A} have provided estimates for the mass-loss rate of a pressure supported globule under the effect of a wind. All their estimates are based on the mass-loss rate of the globule being, due to conservation of momentum, approximately equal to the mass flow rate though the cross-sectional area of the object being ablated, i.e., $\dot{M} = \pi r^{2} \rho_{wind} \nu_{wind}$. Calculating this value for the constant wind used for our simulations gives $\dot{M} = 3 \times 10^{-8}\,$M$_{\sun}\,$yr$^{-1}$ which is around 100 times lower than the value seen in the simulations ($\sim 10^{-6}\,$M$_{\sun}\,$yr$^{-1}$). The flaw in this argument when applied to disks is that each parcel of gas already contains a large amount of momentum due to its rotation. For protoplanetary disks the pressure support is relatively weak so the velocity of the gas is close to Keplerian and thus the escape velocity. This means that the oncoming wind needs to transfer a smaller proportion of its momentum to the gas in the disk so that it becomes unbound. In practice this interaction is fairly complex, particularly for non-face-on orientations and there has been no simple analytical formulae proposed, but it is clear that it is not well described by those for simpler, non-rotating objects.

Photoevaporation by a nearby massive star can also play a role in disk dispersal, and is thought to be the cause of the proplyd objects in the Orion nebula \citep{1997IAUS..182..561H}. \cite{1999ApJ...515..669S} model the effect of an external source of ultraviolet radiation on circumstellar disks, finding good agreement with their models and observations of the Orion proplyds. They provide a crude fit to their results for the mass-loss rate:
\begin{equation}
	\dot{M} \approx 10^{-7} M_{\sun} \mathrm{yr}^{-1} \left( \frac{r_{disk}}{100 \mathrm{au}} \right)^{1 - 1.5},
\end{equation}
at a distance of $0.2$pc from a $\theta^1$ Ori C-like star. \cite{2000ApJ...539..258R} perform a similar analysis and find agreement with the above equation. Calculating this value for a disk of $r_{disk} = 40\,$au gives $\dot{M} = 3 \times 10^{-8}\,$M$_{\sun}\,$yr$^{-1}$. This is $\sim100$x less than the continuous stripping rate from the SNR interaction at a distance of $0.3\,$pc for models \textbf{const00med} and \textbf{const00high}. The lifetime of the const00med and const00high disks to photoevaporation, $t_{\rm life} \approx M_{\rm disk}/\dot{M}$, is thus $3\times10^{4}$ and $3\times10^{5}$\,yrs, respectively. This indicates that photoevaporation can be an important process which may destroy disks before any nearby massive star explodes as a SN. However, not all disks will be exposed to as high a UV flux as occurs in Orion, despite still being subject to the effects of a nearby SN explosion (for instance, a B0V star has an ionizing flux which is $1.5\times$ lower than $\theta^1$ Ori C). In addition, more massive disks than we have considered may survive photoevaporation from Orion-like UV fluxes for several millions of years before being subject to a SN explosion.

Another possible source of ablation for stellar disks is the wind from a nearby massive star. This can last much longer than the SNR. For massive stars ($M \gtrsim 25\,$M$_{\sun}$) The strongest wind occurs during the Wolf-Rayet (WR) phase which lasts $\sim0.3\,$Myr. For an isotropic wind, the ram pressure at a distance, $r$ is
\begin{equation}
	P_{ram} = \frac{\dot{M}}{4 \pi r^2 \nu},
\end{equation}
where $\dot{M}$ is the mass-loss rate of the star and $\nu$ is the velocity of the stellar wind. Typical values for the star's wind during the WR phase are $\dot{M} = 10^{-5}\,$M$_{\sun}\,$yr$^{-1}$ and $2 \times 10^{-8}\,$cm$\,$s$^{-1}$, giving a ram pressure of $1.2 \times 10^{-8}\,$g$\,$cm$^{-1}\,$s$^{-2}$. This is several orders of magnitude lower than appears in any of the three disk masses simulated (see Figure \ref{fig5:radial_gp}) so no direct stripping is possible. While there may be some long term, continuous stripping, we expect stripping from photoevaporation to dominate such periods.

In conclusion, photoevaporation and hydrodynamic ablation are processes which can both affect the disk prior to it being subject to a SN explosion. However, we expect some disks to survive these earlier processes. For those that do, supernova induced stripping can far exceed the photoevaporative mass-loss rate and clearly dominates the mass loss rate during the interaction of the disk with the SNR. This is the scenario which we study in this work.

A final issue concerns the fact that the radiation and winds of nearby massive stars will alter the ambient medium into which the supernova explodes. For instance, a steady wind will produce a density profile which declines as $r^{-2}$. Alternatively, the massive star may undergo an outburst and eject a dense shell of gas (perhaps the most famous example of this is the Homunculus Nebula produced by $\eta$~Carinae). However, unless the mass that the supernova remnant encounters is comparable to the ejecta mass, the SNR will remain in its free-expansion phase and thus its dynamics will be largely unaffected by the surrounding material. In our simulations, the swept up material is a tiny fraction of the ejecta mass at $0.3\,$pc ($\sim 0.015\%$ for the SNR values used here).

\subsection{Planet formation}

As planet formation occurs on timescales much longer than the SNR interaction simulated in this work, the disk is likely to re-establish equilibrium before planets begin to form. As far as planet formation is concerned, the interaction is therefore equivalent to replacing the disk with one of smaller mass.

\subsection{Enrichment via supernova}

Overall very little mass from the SNR flow becomes bound to the disk: $\sim10^{-9}\,$M$_{\sun}$ for face-on disks that survive the stripping (models \textbf{const00med} and \textbf{const00high}) and $\sim10^{-10}\,$M$_{\sun}$ for the edge-on disk that survives (model \textbf{dyn90low}). This is about 10 times less than seen by \cite{2007ApJ...662.1268O}. Some of the difference may be due to the differences in the dimensionality of the simulations (3D in our case, 2D in theirs). However, most of it is likely due to differences in the radiative cooling (which is included in their simulations, while ours are adiabatic). Thus the hot supernova ejecta is more able to cool down and mix with the disk in their simulations. The broad conclusion is the same, however: that the enrichment of protoplanetary disks via supernova ejecta by pure hydrodynamic mixing is too inefficient to explain the abundance of SLRs in the early solar system. The highest \correction{value of $f_{enr}$} seen in our simulations with surviving disks is $\approx 5 \times 10^{-6}$ (in model \textbf{dyn90low} - see Fig.~\ref{fig5:snrbound}), which falls short of the $10^{-4}$ needed using SLR production values from \cite{1995ApJS..101..181W}. Interestingly, this is for our low mass, edge-on case. Low mass disks obviously benefit more from the same amount of ejecta material, but are also generally destroyed more easily. However, we believe that the fact that edge-on disks survive significantly longer than face-on disks allows such disks to intercept more of the SN ejecta, and thus have the highest enrichment fraction. So, at least for the particular parameters chosen here, a low mass edge-on disk can be enriched more than a high mass face-on disk. While the disks here do not meet the enrichment requirements deduced from observations, it may be possible that a much larger disk (and hence larger surface area for collection) placed further away from the supernova (such that its outer parts do not get instantaneously stripped) could be more efficient at absorbing SN ejecta. This is left for the subject of future investigation.

\correction{
We have also only considered gas phase ejecta. It has been suggested \citep[by e.g.][]{2016MNRAS.462.2777G} that dust grains would be able to penetrate and be absorbed more easily into the disk. However, the abundance of dust grains in the SNR is uncertain \citep[see e.g.][]{2007MNRAS.378..973B, 2017MNRAS.465.3309D}. The dust must also survive the reverse shock as the SNR interacts with the surrounding medium \citep{2016A&A...590A..65M} although this may not be significant for the close proximity SNRs required for disk enrichment.
}

Finally we note that the enrichment which we see in our simulations is limited to the surface of the disk. The timescales in our simulations are too short for mixing of the SLRs throughout the disk \citep[][note a mixing time of $10^{4}$\,yr]{2013ApJ...773....5B}, and in any case the resolution is such that we do not adequately resolve the motions responsible for mixing material within the disk.

%__________________________________________________________________

\section{Conclusion}

We have presented three dimensional simulations of the stripping of stellar disks due to the influence of a nearby SNR, using a physically motivated dynamic flow. We have also investigated the effect of varying the inclination angle and disk mass.While a number of other processes (e.g. photoevaporation and ablation by the radiation field and wind from a nearby massive star) can cause significant stripping, or even complete destruction of the disk, prior to interaction with a SNR, this is not always the case. We therefore assume in our investigation that at least some part of the disk is present when a nearby massive star explodes. 

Good agreement is found with the analytical predications of \cite{2000ApJ...538L.151C}. However, this only accounts for part of the stripping as Kelvin-Helmholtz instabilities can cause additional material to be ablated. In the initial, instantaneous stripping phase, a flow at the peak ram pressure can strip $90\%$ of a low mass disk ($M_d \sim 0.1 M_J$) and $30\%$ of a medium mass disk ($M_d \sim 1.0 M_J$) on timescales of $10$-$100\,$yrs (less than one outer rotation period). High mass disks ($M_d \sim 10 M_J$) are largely unaffected by instantaneous stripping.

During continuous, longer-term, ablation disks lose mass at a rate of $\sim10^{-6}\,$M$_{\sun}\,$yr$^{-1}$. This value decreases with time as the SNR passes and the flow weakens, but is several orders of magnitude greater than the mass-loss rate due to photoevaporation or stellar wind ablation and will therefore dominate the disk's mass-loss rate during this time.

We find that the inclination angle only has a large effect on the evolution when the disk is close to edge on (similarly to previous findings from simulations of disk galaxies). When the ram pressure is large compared to the gravitational pressure in the disk, low inclination angle disks are deformed to the point that their evolution converges to that of a face-on (zero degree inclination) disk. In contrast, our edge-on disks show a much steadier rate of mass-loss (instantaneous stripping is much reduced due to the lower cross-section) and can survive significantly longer than their face-on counterparts. Amongst the low-mass disks simulated, only the edge-on disk survived interaction with the SNR (retaining almost 60\% of its mass).

The stripping of inclined disks can be quite asymmetrical, and the direction of the stripped tail may not line up with the direction of the flow (also like disk galaxies). However, unlike disk galaxies, the flow may die down before the asymmetries have disappeared. For the SNR parameters chosen the interaction is very short, lasting only a couple of hundred years. This means that stellar disks are unlikely to be observed during this period, but seeing an asymmetric disk may be evidence that it underwent this type of ablation in its past. For disks more distant from the supernova the duration of significant interaction will be longer. 

Amongst all of our simulations, the highest ejecta enrichment fraction is $5 \times 10^{-6}$, which is too low to explain the presence of SLRs in the early solar system. The highest enrichment is seen in a low mass edge-on disk, suggesting the ideal case for enrichment is a low mass edge-on disk (that would be destroyed if placed face-on) rather than a face-on disk which one might naively assume.

\section*{Acknowledgements}

The calculations for this paper were performed on the DiRAC 1 Facility jointly funded by STFC, the Large Facilities Capital Fund of BIS and the University of Leeds.

\bibliographystyle{mnras}
\bibliography{pdisk_ablate}

\bsp
\label{lastpage}
\end{document}